\documentclass[12pt]{iopart}
\usepackage{graphicx}
\usepackage{amssymb}
\usepackage{float}
\usepackage{eqnarray}
\usepackage{lineno}
\usepackage{caption}
\usepackage[utf8x]{inputenc}
\usepackage{cite}
\usepackage{subcaption}
\usepackage{graphicx}
\usepackage[ 
    colorlinks=true,
    pdfborder={0 0 0}, linkcolor=blue,
    filecolor=magenta,      
    urlcolor=cyan
]{hyperref}

\begin{document}
%%   round  -  round parentheses are used (default)
%%   square -  square brackets are used   [option]
%%   curly  -  curly braces are used      {option}
%%   angle  -  angle brackets are used    <option>
%%   semicolon  -  multiple citations separated by semi-colon
%%   colon  - same as semicolon, an earlier confusion
%%   comma  -  separated by comma
%%   numbers-  selects numerical citations
%%   super  -  numerical citations as superscripts
%%   sort   -  sorts multiple citations according to order in ref. list
%%   sort&compress   -  like sort, but also compresses numerical citations
%%   compress - compresses without sorting
%%
%% \biboptions{comma,round}

% \biboptions{}

%\journal{Journal Name}
%\tableofcontents

%\begin{frontmatter}

\title{Reducing Scattered Light in LIGO's Third Observing Run.}



%\author{Siddharth Soni}

%\address{Louisiana State University, Louisiana, USA.}

% \author{ S. Soni$^1$, C. Austin$^1$, A. Effler$^2$,  R. Schofield$^{3,4}$, G. Gonz\'alez$^1$, V. Frolov$^2$, J. Driggers$^4$, A. Pele$^2$,  A. Urban$^1$, G. Valdes$^1$ }\address{$^1$ Department of Physics and Astronomy, Louisiana State University, Baton Rouge, LA, 70803, USA}\address{$^2$ LIGO Livingston Observatory, Livingston, LA 70754, USA}\address{$^3$ Department of Physics, University of Oregon, Eugene, OR 97403, USA}\address{$^4$ LIGO Hanford Observatory, Richland, WA, 99352, USA} \ead{ssoni3@lsu.edu}

\author{%
S~Soni$^{5}$,  %siddharth.soni
C~Austin$^{5}$,  %corey.austin
A~Effler$^{2}$,  %anamaria.effler
R~M~S~Schofield$^{7}$,  %robert.schofield
G~Gonz\'alez$^5$, %gaby.gonzalez
V~V~Frolov$^{2}$,  %valery.frolov
J~C~Driggers$^{10}$,  %jenne.driggers
A~Pele$^{2}$,  %arnaud.pele
A~L~Urban$^{5}$,  %alexander.urban
G~Valdes$^{5}$
% ...
\\
{(The LIGO Scientific Collaboration)}%
}%
\par\medskip
% ...
\address {$^{2}$LIGO Livingston Observatory, Livingston, LA 70754, USA }
% ...
\address {$^{5}$Louisiana State University, Baton Rouge, LA 70803, USA }
% ...
\address {$^{7}$University of Oregon, Eugene, OR 97403, USA }
% ...
\address {$^{10}$LIGO Hanford Observatory, Richland, WA 99352, USA }
% ...
\begin{abstract}

Noise due to scattered light has been a frequent disturbance in the Advanced LIGO gravitational wave detectors, hindering the detection of gravitational waves. The non stationary scatter noise caused by low frequency motion can be recognized as arches in the time-frequency plane of the gravitational wave channel. In this paper, we characterize the scattering noise for LIGO and Virgo's third observing run O3 from April, 2019 to March, 2020. We find at least two different populations of scattering noise and we investigate the multiple origins of one of them as well as its mitigation. We find that relative motion between two specific surfaces is strongly correlated with the presence of scattered light and we implement a technique to reduce this motion.  We also present an algorithm using a witness channel to identify the times this noise can be present in the detector. 
%To identify the presence of the noise, we use data from sensors that monitor light transmitted through the end test masses of the detector.

\end{abstract}

%\begin{keyword}
%Science \sep Publication %\sep Complicated
%%% keywords here, in the %form: keyword \sep %keyword
%
%%% MSC codes here, in the %form: \MSC code \sep code
%%% or \MSC[2008] code %\sep code (2000 is the %default)
%
%\end{keyword}

%\end{frontmatter}

%\linenumbers

\section{Introduction}\label{introduction}
%\label{S:1}

\begin{figure}[h]
    \centering
    \includegraphics[width=12cm]{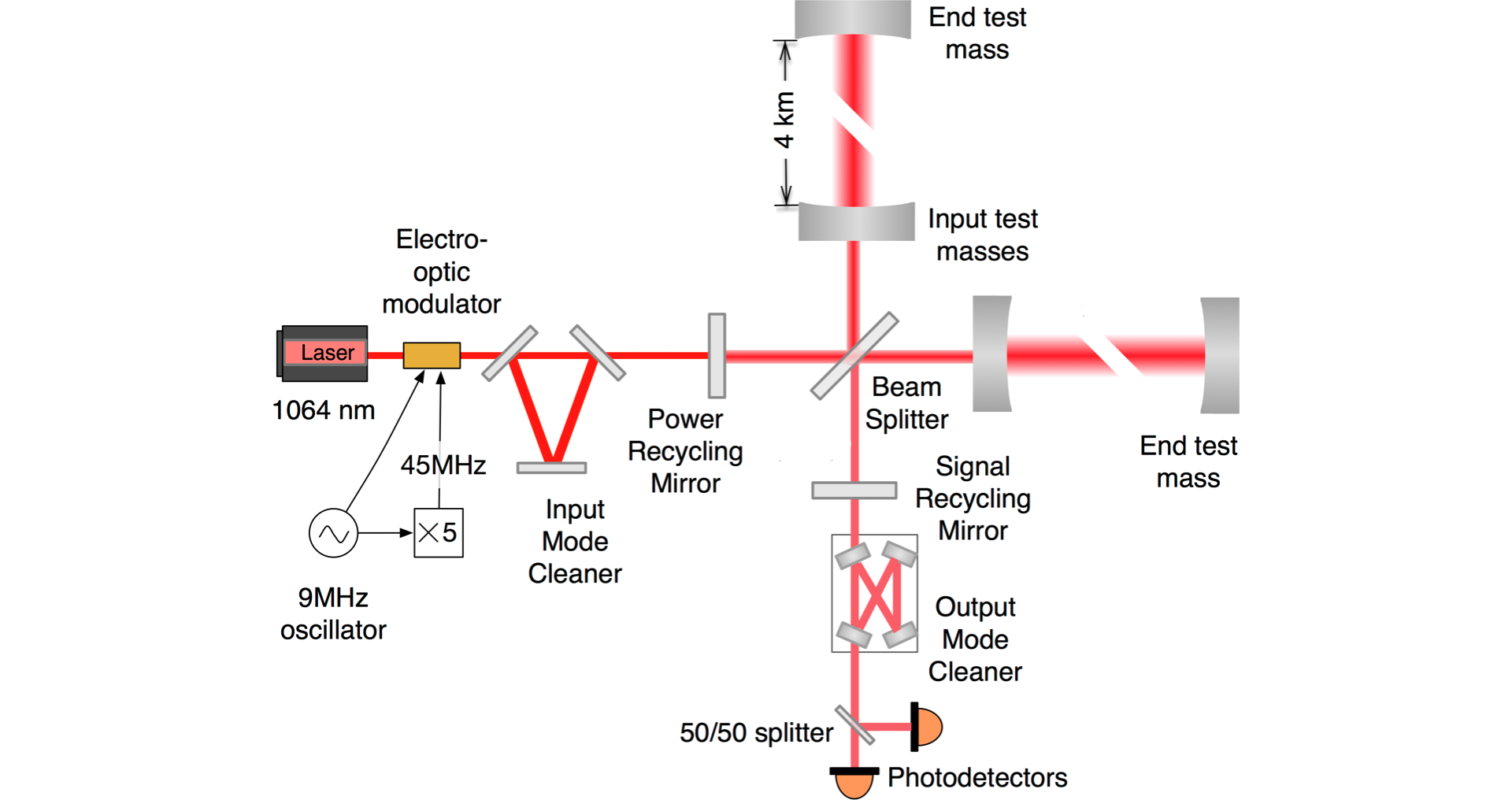}
    \caption{A schematic of aLIGO detectors taken from {~\cite{Martynov_2016}} }
    \label{fig:schematic}
\end{figure}

%Maecenas \cite{Smith:2012qr}% %fermentum \cite{Smith:2013jd}% 
The LIGO  gravitational-wave observatories located at Hanford, Washington (LHO), and Livingston, Louisiana (LLO) in the USA \cite{TheLIGOScientific:2014jea}, along with the Virgo detector in Cascina, Italy~\cite{Acernese_2014}, and the GEO 600 detector in Germany~\cite{Dooley_2016} are a part of a worldwide network of gravitational-wave detectors. 
A schematic of the LIGO detectors is shown in Fig. \ref{fig:schematic}.
Each LIGO detector is a dual-recycled Fabry-Perot Michelson interferometer with 4 km arms. The detector acts as a transducer for strain, converting phase shift due to a gravitational wave into a signal that can be measured on a photo-diode. The output signal at the photodetectors is calibrated to an equivalent strain signal h(t) \cite{Abbott_2017,Viets_2018}.

The first two observing runs, in September 2015-January 2016 and November 2016-August 2017, resulted in spectacular discoveries, including signals from the merger of 10 pairs of black holes and one from a merger of neutrons stars~\cite{Abbott_2019,Abbott_2016,TheLIGOScientific:2017qsa}. The third observing run began on April 1, 2019, and ended on March 27, 2020. During this run, plausible gravitational wave sources were shared as public alerts, averaging one a week ~\cite{gracedb}. The average binary neutron star (BNS) range at Livingston and Hanford, during O3, is approximately 130 Mpc and 110 Mpc respectively. In October 2019, a month-long working break separated the first and second half of O3, called O3a and O3b respectively. An increase in the laser input power and squeezed light injection contributed to the increase in range from O2 to O3~\cite{prospect}.

Noise from several different sources limits the sensitivity of the strain data at different frequencies. 
While quantum shot noise is dominant at frequencies above 300 Hz~\cite{tjthesis}, ground motion is the major source of noise below 10 Hz which can affect the higher frequency band of the detectors. In addition to quantum noise, suspension noise and seismic noise, short duration noise transients also known as ``glitches'' can affect the sensitivity of the detectors \cite{transientpaper}. These noise transients can be due to high ground motion, electronic malfunction or other reasons, not well-understood \cite{Nuttall_2018,noisepaper,Cabero_2019}. At aLIGO they are detected by the Omicron algorithm. Each of these glitches, also known as omicron trigger, is annotated with some parameters such as event time, frequency, signal to noise ratio (SNR) ~\cite{omicron_florent,robinet2020omicron,McIver:2015pms}. In this paper, we focus on the type of noise transient due to scattered light which impacts the gravitational strain in 10 Hz to 120 Hz frequency band. Signals from the merger of very massive binary black holes may fall into this frequency region \cite{GW190521Adiscovery}. Thus, identifying and reducing the amount of scattering is imperative for increased confidence in detecting such signals and overall enhanced detector sensitivity. 
 
The paper is composed as follows. In Sec. \ref{hardware} we give an overview of the sensors and instruments at the LIGO end stations. Next in Sec. \ref{scatteringnoise}, we discuss scattering noise and mathematically define the phase and amplitude component of the noise. Scattering noise during O3 is addressed in Sec. \ref{scatino3}. In Sec. \ref{m0r0_scattering} we provide a mathematical treatment of slow scattering, a sub-population of the light scattering and discuss the source of the noise. In Sec. \ref{rzero} we develop a method to mitigate the slow scattering noise discussed in Sec. \ref{m0r0_scattering}. Sec. \ref{transmon_scattering} introduces another source of slow scattering observed during O3. In Sec. \ref{transmonwitness} we discuss a method to identify scattering times. Finally we summarize the paper in Sec. \ref{summary}. 

\section{Hardware and sensors at LIGO end stations}\label{hardware} 
\begin{figure}
    \centering
    \includegraphics[width=\textwidth]{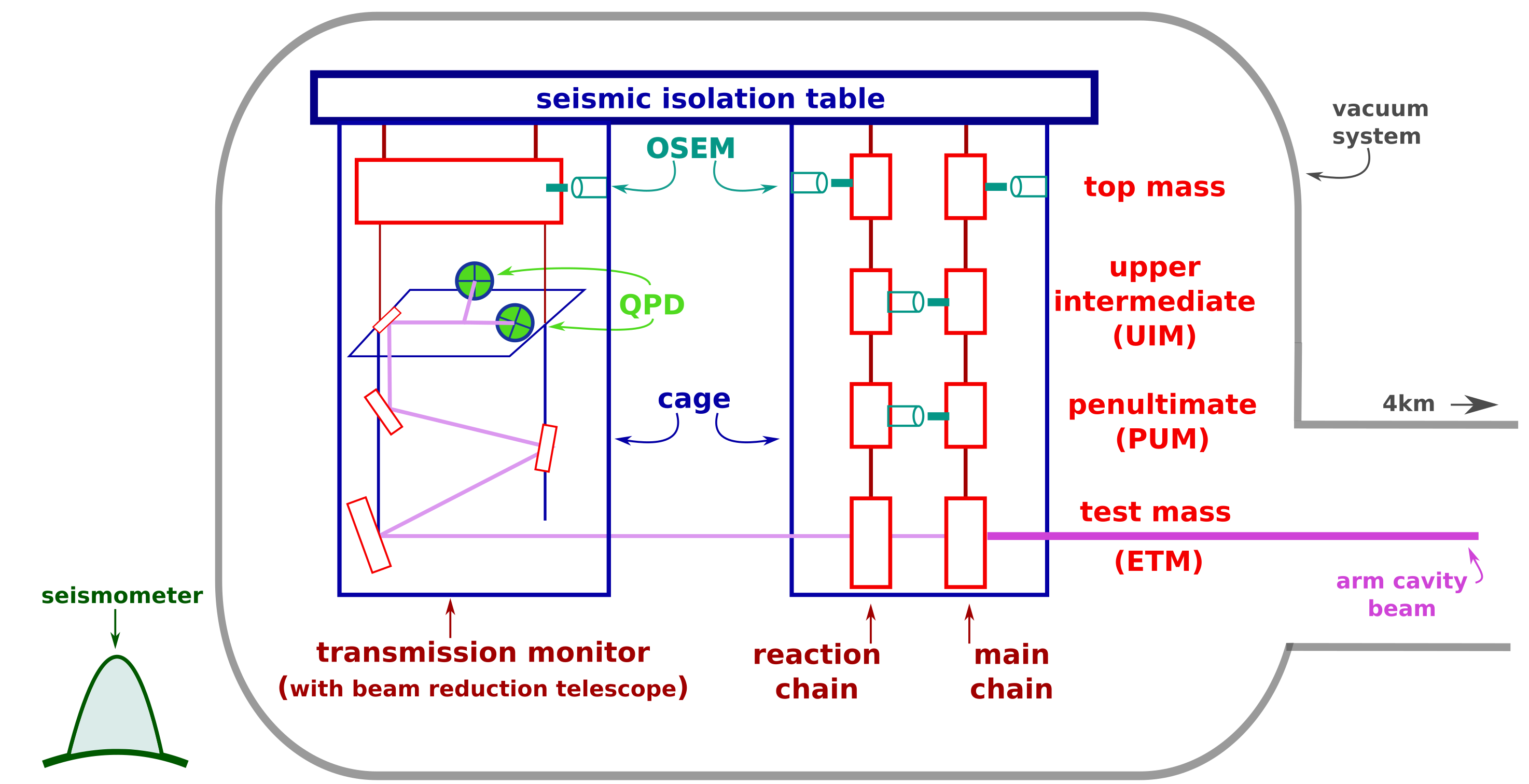}
    \caption{The schematic of end station housing shows the seismic isolation system, quadruple suspension, transmission monitor system (TMS), and the seismometer used to measure ground vibrations. The optical shadow sensor and magnetic actuators (OSEMs) in the quadruple suspension cage measure relative motion and produce a force on the main and reaction chain. About 4 ppm of the light power in the arms is transmitted through the end test mass. $95$\% of this 4 ppm is dumped and the remaining $5$\% is split onto the two-quadrant photo-diodes (QPDs) in the TMS. These QPD's are used to check for any misalignment of the beam on the end test mass. }
    \label{fig:end_station}
\end{figure}

The site of light scattering discussed in this paper, is the set of mirrors located in the end station housing. This section describes these mirrors and other hardware, crucial for an understanding of light scattering.
A schematic of the LIGO end station housing one end test mass \cite{Pinard:17} of the interferometer arm is shown in Fig. \ref{fig:end_station}. Each station has a reference seismometer on the floor to monitor ambient seismic noise. The vacuum system has at the end of the 4km arms large vacuum chambers housing the test masses. The test mass and a transmitted light monitoring table are suspended from a two-stage seismic isolation table \cite{Matichard_2015}. The test mass has a double chain, four-layered suspension for additional isolation \cite{Aston_2012}. The quadruple suspension chain behind the quadruple test mass suspension is called the reaction chain. There is additional hardware around the suspension, most importantly the ``cage": a structure hard bolted to the seismic isolation table which serves both as a reference and as a safety measure for protecting against large motions that could damage the suspension.

The suspension for the test mass uses optical shadow sensor and magnetic actuators (OSEM)  at the top three stages as well as an electrostatic drive (ESD) at the test mass stage. The OSEMs at the top stages measure relative motion and can apply a force between the cage and the two chains. For the upper-intermediate mass (UIM) and penultimate (PUM) mass stages, the OSEMs measure and actuate in-between the two chains. 

The transmission monitor is a double suspension with OSEMs at the top stage, measuring motion with respect to its cage \cite{Barsotti_2010}. It houses a telescope to reduce the large beam size coming from the arm. The LIGO end test masses have a transmission of approximately 4ppm, so for an arm power of 250kW as in O3a for the Livingston detector, the transmitted beam would be about 1W. $95\%$ of this light is dumped and $5\%$ is split about equally between two quadrant photodiodes.

In order to keep the optical cavities in the arms on resonance, LIGO uses the interferometric signal at the detector output to sense the difference in the arm length. It then feeds back that signal to one of the end test masses, at different stages of the quadruple suspension, with a bandwidth of about 60 Hz \cite{Martynov_2016}. As a result, the test mass chain moves much more relative to the local surfaces, since it has to ``account" for the motion of all other test masses. The main pendulum frequency of the quadruple suspension is around 0.45 Hz which means that for the force applied at frequencies below 0.45 Hz at upper stages, the entire chain will move together. The dominant ground motion to be fed back is indeed below this frequency. Furthermore, because most of the force is applied in between the chains, the test mass chain moves twice as much relative to the reaction chain. The test mass motion relative to other surfaces like the transmission monitor, the cage or the vacuum chamber walls is half of the motion measured by the OSEM at the UIM or PUM level, below 0.45 Hz.

Each LIGO detector is equipped with several hundreds of auxiliary sensors, used in the feedback control system and in the environmental monitoring system. Many of these channels are not sensitive to differential arm length and are used to identify various environmental and physical couplings to the detector.
Ground motion in various frequency bands, for example, is measured by seismometers located at end stations and corner station. These seismometers record ground motion in X, Y and Z direction in the frequency range from 0.03 Hz to 30 Hz. Earthquakes shake the ground in $0.03-0.1$ Hz band while seismic noise due to trains and human activity near the site, also known as anthropogenic noise, shows up in the $1-3$ Hz region. Ocean waves and sea storms produce seismic waves with frequencies ranging from $0.03-0.5$ Hz, also known as microseisms. While the secondary, and dominant, microseism peak is typically measured around $0.15$ Hz \cite{Cessaro1994SourcesOP,Effler_2015}, it varies in frequencies and was strongest at $0.13$ Hz for this analysis. As we discuss later, the output of these sensors is used to look for correlations with noise transients in the strain data.

\section{Scattering Noise}\label{scatteringnoise}
Tiny imperfections on the surfaces of test mass mirrors in the interferometer cause a small amount of light to scatter out of the main beam. This scattered light can then reflect from surfaces that have large relative motion relative to the test mass such as the chamber walls and then back to the test mass. Upon recombining with the main beam, the scattered light introduces noise in the gravitational wave data. The amplitude of the noise depends on how much light recombines with the main beam, and the upper frequency depends on the relative motion.

The motion of the scatterer introduces an additional phase in the field reflected from its surface. Consider a small fraction $A$ of the total field that gets scattered back to the main beam, from a scattering surface located behind the end test mass (ETM). This light will acquire an additional phase  due to the path length modulation caused by the relative motion between the end test mass and the scatterer. The resulting phase noise can be formulated as:
\begin{eqnarray}
      h_{ph}(f) & = & A\frac{{\lambda}}{8{\pi}L}\mathcal{F}\left[\sin{\delta{\phi}(t)}\right] \label{eq:1}
\end{eqnarray}

\begin{eqnarray}
    {\phi}(t) & = {\phi}_{0} + {\delta\phi_{sc}(t)} = & \frac{4{\pi}}{{\lambda}}\left|{ x_{0} + {\delta} x_{sc}(t)}\right| \label{eq:2}
\end{eqnarray}
    
${\lambda}$ is the laser wavelength, $x_{0}$ is static path that corresponds to the static phase $\phi_{0}$ while  $\delta x_{sc}$ is the time-dependent displacement of the scattering surface which gives rise to additional phase $\delta \phi_{sc}(t)$, 
% which is added to the static path $x_{0}$, $L$ is the length of arms (4 kms), 
$\mathcal{F}$ is the Fourier transform ~\cite{Accadia:2010zzb,Canuel:13,vajente_chapter}. 
\par
The build up of this phase shifted field in the arms by the factor ${\Gamma}$ pushes on the mirrors resulting in radiation pressure noise. This radiation component of the noise can be expressed as:

\begin{equation}
    h_{rad}(f) = A\frac{2{\Gamma P}}{Mc}\frac{2}{\Omega^2 - \omega^2} \mathcal{F}\left[\cos{\delta{\phi}(t)}\right] \label{rad_eqn}
\end{equation}
${\Gamma}$ $= 13.58$ here is cavity signal gain, M $= 40$ kg is the mass of mirrors, P is the power in the arms, $c$ is speed of light and ${\Omega}$ is the suspension eigenfrequency~\cite{Ottaway:12}.
\begin{figure}[h]
    \centering
    \includegraphics[width=10cm]{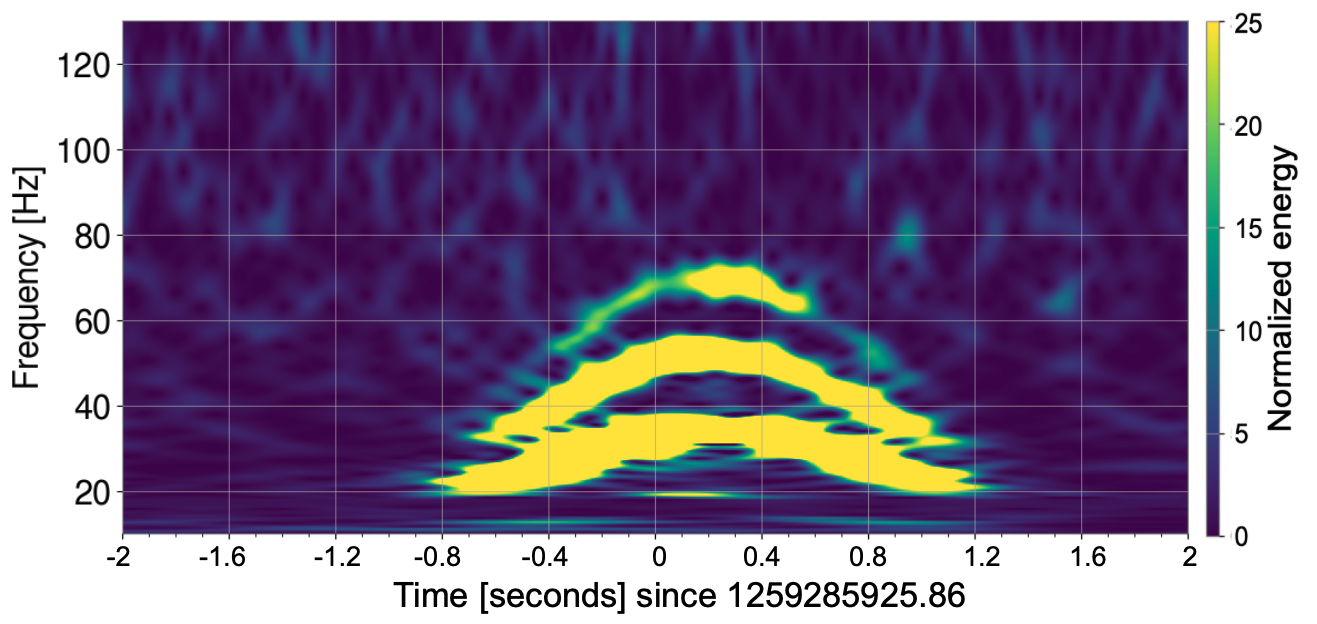}
    \caption{Scattering noise shows up as arches in the time-frequency spectrograms also known as Q-scan of the primary gravitational wave channel as processed by \textbf{gwdetchar-omega}\cite{Chatterji:2004qg,gwdetchar}. The stacked arches suggests a scatter path that involves multiple reflections of stray light between the test mass and the scatterer.}
    \label{fig:scat}
\end{figure}

Scattering noise can be recognized as arches in the Q-scan  as shown in Fig. \ref{fig:scat}. The peak frequency of these arches tells us the number of fringes per second and can be related to the velocity of the scatterer, $v_{sc}$ with the following relation:
\begin{equation}
    f_{\rm fringe}(t) = \left|\frac{2nv_{sc}(t)}{\lambda}\right|  \label{eq:3}
\end{equation}
where $f_{\rm fringe}(t)$ is the fringe frequency and $n$ is the number of times stray light gets reflected back and forth between the test mass and the scatterer before it joins the main beam. This equation can be derived by calculating the rate of change of phase from Eq. \ref{eq:2}.  We can also look at the spacing of the peaks in the time domain to give us half the period of the scattering surface.

\section{Scattering in O3}\label{scatino3}
The BNS range defined as the radius of the orientation-averaged spacetime volume assuming a
matched-filter detection signal-to-noise ratio threshold of 8, increased to 110-140 Mpc in O3, from 80-100 Mpc in O2~\cite{prospect} Due to this improvement in sensitivity and the enhanced stability of the interferometer which allows the detector to stay operational during high microseismic activity, some of the transients of similar origin in O2 and O3 now surface with higher signal to noise ratio (SNR). 
Consequently, we see a lot more scattering arches in O3. An interesting feature of scattering in O3 is the presence of two different populations of scattering triggers, so-called the ``slow'' scattering and the ``fast'' scattering at both LLO and LHO. The glitch morphology of the slow scattering is the more familiar arch in Q-scan as shown in Fig. \ref{fig:scat}. The fast scattering triggers are more localized in time and occur with lower SNR compared to slow scattering. Table \ref{tab:fastslowtab} compares different characteristics of slow and fast scattering triggers as classified by a noise classification tool GravitySpy discussed in Sec. \ref{transmonwitness}.

\begin{figure}[h]
\captionsetup[subfigure]{font=scriptsize,labelfont=scriptsize}
   \centering
    \begin{subfigure}[b]{0.45\textwidth}
        \centering
         \includegraphics[width= \textwidth]{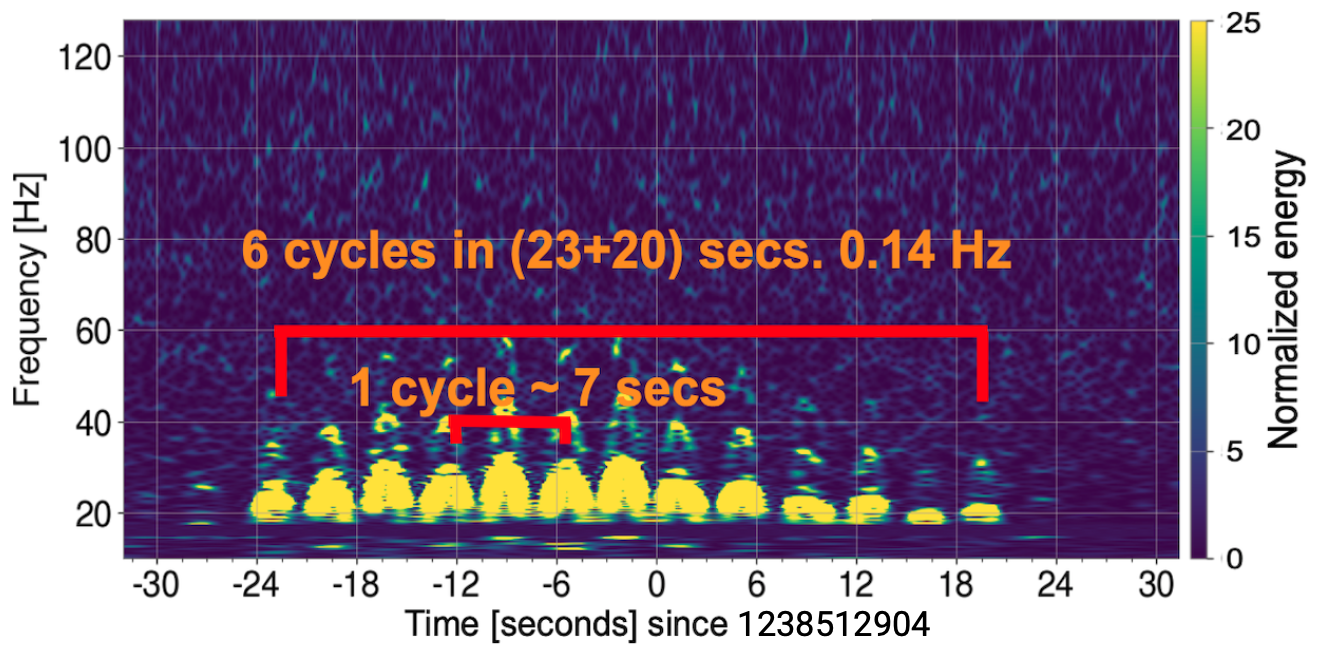}
         \caption{Slow scattering arches.}
         \label{fig:slowscat}
    \end{subfigure}
    \hfill
    \begin{subfigure}[b]{0.47\textwidth}
        \centering
         \includegraphics[width =\textwidth]{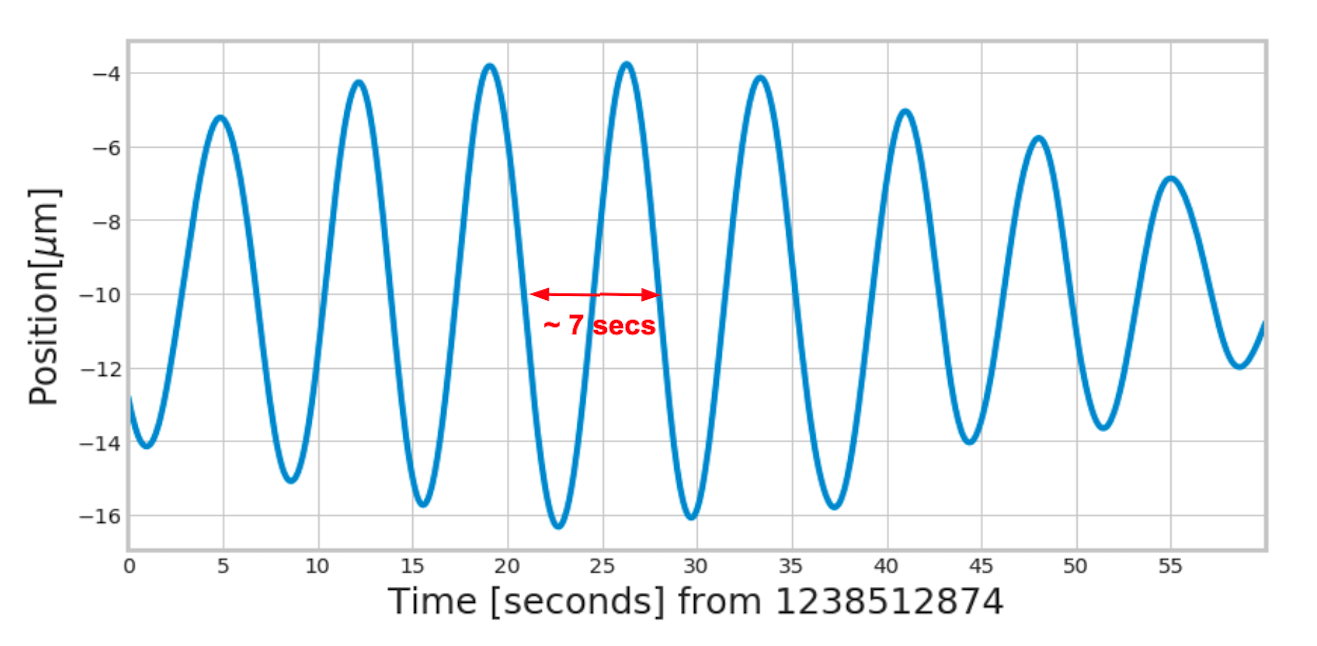}
         \caption{PUM stage motion during light scattering.}
         \label{fig:timeseriessus}

    \end{subfigure}
    \caption{The left image shows Q-scan of  slow scattering in the primary gravitational wave channel. This is an unusual number of arches due to extremely high ground motion in the $0.03$ - $0.1$ Hz (earthquake) and $0.1$ - $0.3$  Hz (microseism) bands on April 5, 2019. The right image shows the PUM stage motion between the chains for the same time period. As mentioned in Sec. \ref{hardware}, the PUM stage OSEM records twice the displacement of the main chain.}
    \label{fig:slowscat_sus}
    
\end{figure}

\subsection{Slow Scattering.}
In time-frequency spectrograms, slow scattering triggers resemble an arch.  Fig. \ref{fig:slowscat}. shows slow scattering in h(t) for a day with large ground motion. As shown in this figure, the scatterer has a period of $\sim$ 7 seconds, and from this we can derive the scattering surface is moving with a frequency close to 0.13 Hz  which corresponds to one of the peaks in microseism ~\cite{alogsid}.
Slow scattering is  dominant during high ground motion in $0.03$ - $0.1$ Hz (earthquake) and $0.1$ - $0.3$  Hz (microseism) band. These arches reach high frequencies during larger ground motion and so it is more visible above the quiet background noise in the differential arm cavity (DARM). During O3, it was particularly strong on Dec 1, 2019 and January 6, 2020 due to high levels of ground motion on both of these days at LLO.  Depending on the ground motion, slow scattering creates ``scatter shelves'' in the frequency band 10 Hz to 120 Hz in h(t) spectra.

\subsection{Fast Scattering.}
\begin{figure}
\captionsetup[subfigure]{font=scriptsize,labelfont=scriptsize}
   \begin{minipage}{0.45\textwidth}
   \begin{subfigure}[b]{\textwidth}
        \centering
         \includegraphics[width=\textwidth,height=4cm]{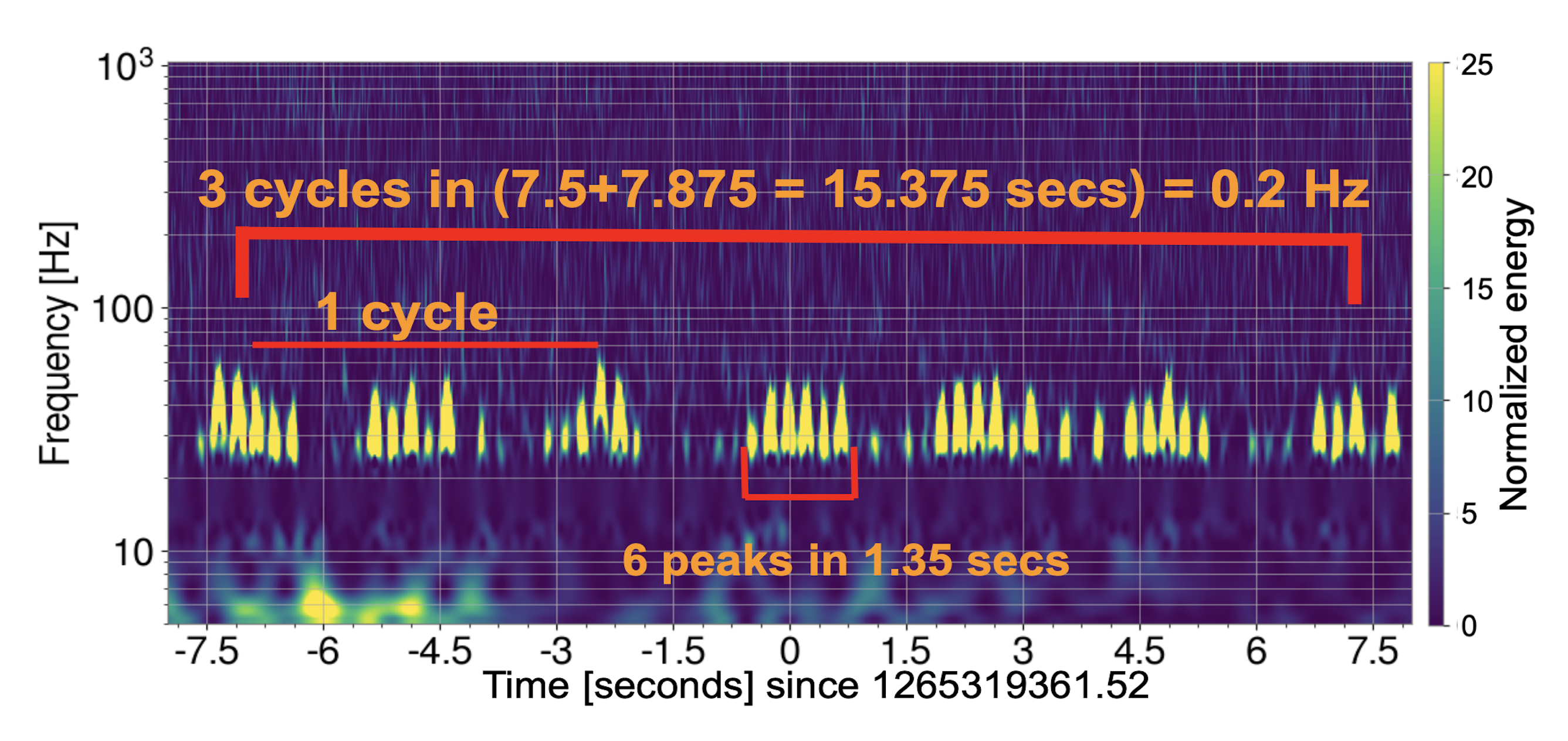}
         \caption{Fast scattering triggers.}
         \label{fig:fastscat}
    \end{subfigure}
    \end{minipage}
    \begin{minipage}{0.45\textwidth}
    \begin{subfigure}{\textwidth}
    \includegraphics[width=\textwidth,height=4cm]{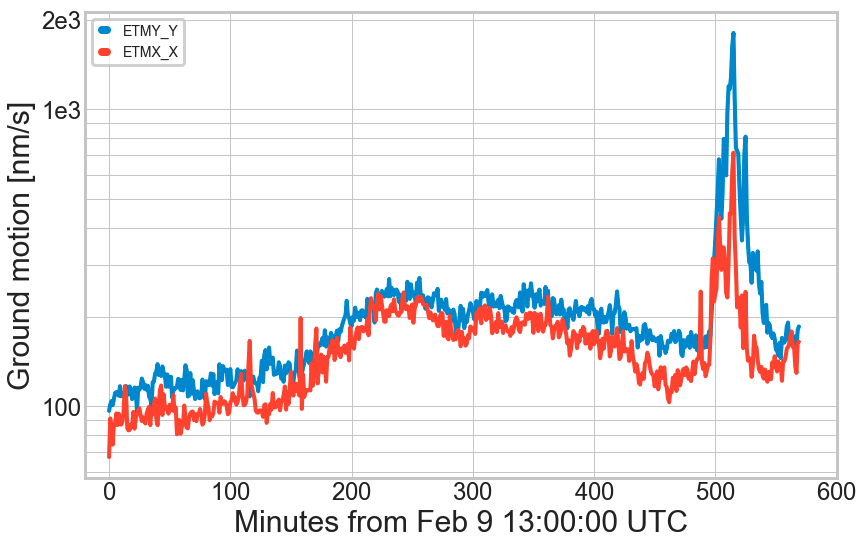}
    \caption{Anthropogenic band ground motion.} \label{subfig:anthro}
    \end{subfigure}

    \vspace*{0.6cm}
    \begin{subfigure}{\textwidth}
    \includegraphics[width=\textwidth,height=4cm]{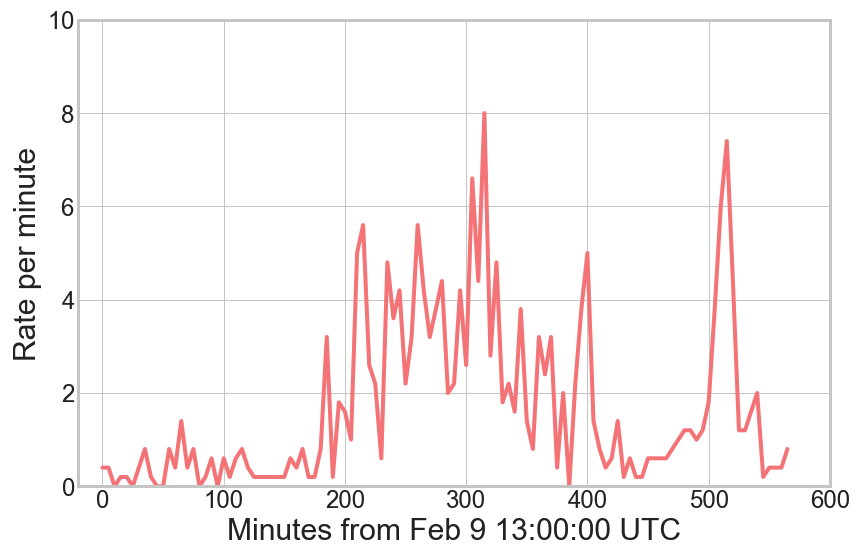}
    \caption{Omicron trigger rate.} \label{subfig:fast_scat_rate}
    \end{subfigure}
    \end{minipage}
    \caption{The left image shows Q-scan of fast scattering triggers in primary gravitational wave channel at LLO. Each ``arch'' contains multiple sub-arches. Fast scattering triggers correlate well with high ground motion in the $1$ - $5$ Hz (anthropogenic) band. The top right image shows the ground motion in the anthropogenic band at X and Y end of the detector at LLO on Feb, 9, 2020. The bottom right image shows the rate of omicron triggers in the frequency band 10 to 50 Hz for the same duration. }
    \label{fig:Fast_scattering}

\end{figure}

The fast scattering triggers shown in Fig. \ref{fig:fastscat} occur with a frequency $\sim$ 4 Hz ~\cite{alogjosh}. This population of scattering correlates with increased ground motion in the  $1$ - $5$ Hz (anthropogenic) and $0.1$ - $0.3$ Hz (microseism) band. It is normally higher in the daytime during the weekdays. Human activity near the site and trains passing on the track near the Y end station at LLO shakes the ground in these frequency bands. 
%June 3, 2019, a day with very high ground motion in the 1-3 Hz band, registered close to 1000 triggers that resembled fast scattering. 
These triggers affect the h(t) sensitivity in the band between 10 and 50 Hz.

\begin{table}[h]
    \centering
    \begin{tabular}{c|c|c}
    \hline
      \textbf{Properties}   &  \textbf{Slow scattering triggers} & 
      \textbf{Fast scattering triggers}\\
      \hline
       Frequency of arches  & below 0.2 Hz & between 2 and 5 Hz \\
      \hline
     % Median Peak Frequency & 20.6 Hz & 34.6 Hz \\
     % \hline
      Median SNR & 37.6 & 11.0 \\
      \hline
      Median duration & 3.2 sec & 1.3 sec \\
      \hline
      \% of total scattering & 40.27 \% & 59.73 \% \\
       
    \hline
    \end{tabular}
    \caption{Comparison of slow and fast scattering triggers in O3a at LLO as identified by GravitySpy above a confidence of $90 \%$ and SNR above 10. Frequency of arches relates to the frequency of the ground motion band active during the noise. Slow scattering is dominant during ground motion in the microseism band ($0.1-0.3$) Hz, whereas Fast scattering is more common during high ground motion in the anthropogenic band ($1-5$) Hz. This comparison is shown for LLO since at LHO, fast scattering amounts to only 1.6\% of the total scattering observed. This is primarily due to difference in ground motion in anthropogenic band between the two sites. }
    \label{tab:fastslowtab}
\end{table}{}

\begin{figure}[h]
\captionsetup[subfigure]{font=scriptsize,labelfont=scriptsize}
   \centering
    \begin{subfigure}[b]{0.45\textwidth}
        \centering
         \includegraphics[width= 6cm]{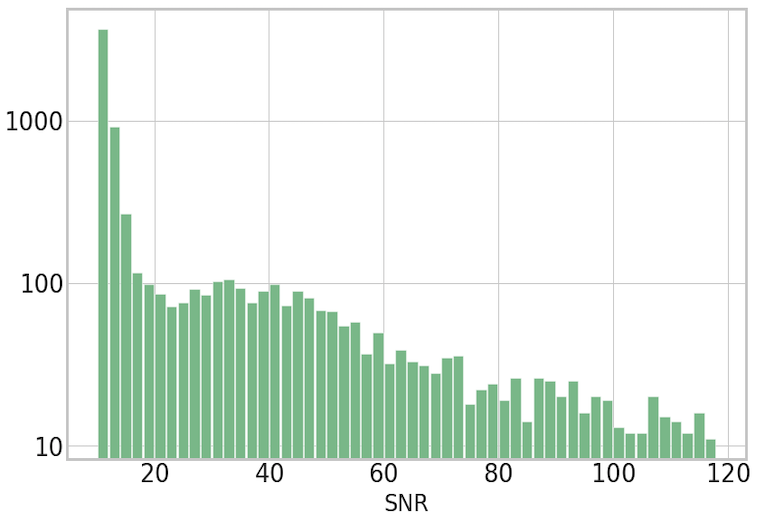}
         \caption{SNR distribution for scattering in O3a.}
         \label{fig:histsnr}
    \end{subfigure}
    \hfill
    \begin{subfigure}[b]{0.5\textwidth}
        \centering
         \includegraphics[width =6cm]{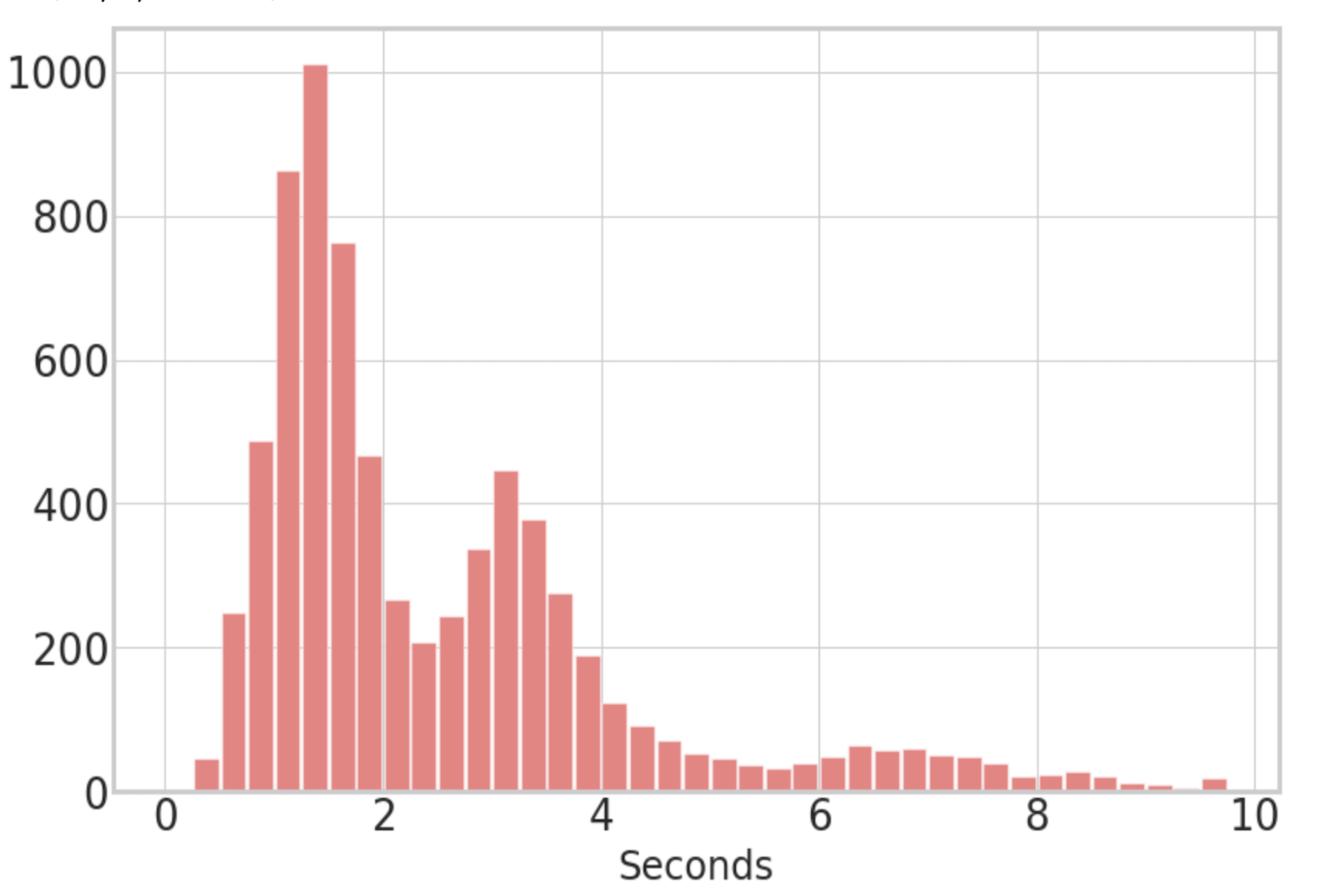}
         \caption{ Duration distribution of scattering in O3a.}
         \label{fig:histdur}

    \end{subfigure}
    \caption{The SNR and duration histograms for total scattering in O3a at LLO reveal the presence of two populations the fast and slow scattering. Slow scattering tends to be louder and long duration. It should be emphasized that the third bump in the duration plot between 6 sec and 8 sec does not correspond to any third population, but rather to extended slow scattering. This is common when high ground motion persists long enough without the interferometer breaking lock.}
    \label{fig:snrdur}
    
\end{figure}

 The striking differences in the glitch morphology, SNR, and the duration for slow and fast scattering triggers suggest that they are due to different noise coupling mechanisms. Fig. \ref{fig:snrdur} shows the SNR and duration of total scattering triggers in O3a. Both distributions reveal the presence of more than one population of scattering triggers. We concentrate in this paper on slow scattering triggers and will not investigate fast scattering further. In the next section, we provide a detailed description of slow scattering noise.

\section{Slow scaterring noise coupling mechanism}\label{m0r0_scattering}
For most of the first three aLIGO observing runs, slow scattering noise occasionally polluted the $h(t)$ spectra during periods of high ground motion in 0.1-0.3 Hz band. The characteristic scattering arches indicated that there were wavelength-scale or larger modulations of the scattering path.
As described in Sec. \ref{hardware} an external drive is applied to the test mass chain to keep the optical cavities on resonance. Because the ground moves differently at the ends of the 4 km long cavities, this drive can lead to micron-scale relative motion between the end test mass (ETM) and other objects in its vicinity, making this region a good candidate for the source of scattering arches.

Several clues pointed specifically towards a scattering path involving the annular end reaction mass (AERM): first, the presence of several harmonics of the arches or scattering shelves, indicated that significant fractions of the light traversed the scattered light path more than once. This eliminated several potential paths, such as to light baffles or enclosure walls, because imperfect reflections on these other paths would likely cause the loss of much more than $90\%$ of the power in each successive round trip. Second, the observation of similar harmonic series of arches at both LHO and LLO suggested that the noise was not due to an improbable alignment. And, third, micron-scale relative motions were recognized between upper stages of the test mass and reaction mass chains, suggesting that the scattering surfaces were likely between the chains. 

 \begin{figure}[h]
    \centering
    \includegraphics[width=10cm]{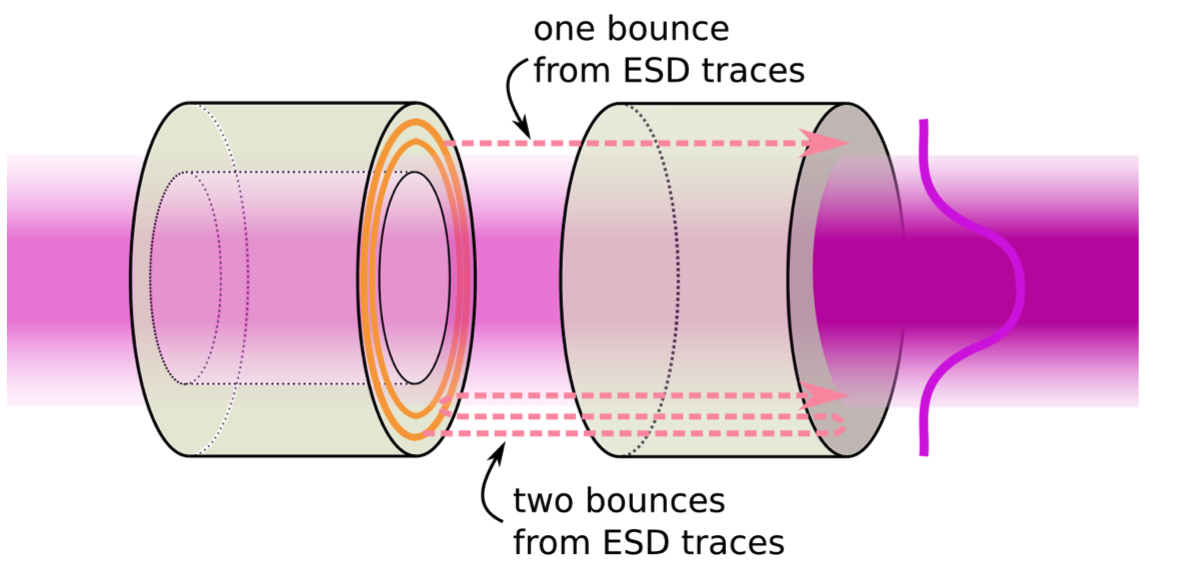}
    \caption{The reaction chain pushes on the main chain  to keep the detector on resonance. Control is applied via coil actuators as shown in Fig. \ref{fig:end_station} and electrostatic drive as shown here. This fluctuates the distance between the AERM shown on the left and ETM shown on the right. A part of the light reflected back from the gold electrostatic drive joins the main beam in the arm with an additional phase. The changing difference in the phase between the two beams introduces noise by causing light modulation at the gravitational wave detection port. Multiple bounces between the ESD trace and the end test mass show up as multiple arches in the h(t) spectrograms.}
    \label{fig:mirror_diag}
\end{figure}
\begin{figure}[h]
\captionsetup[subfigure]{font=scriptsize,labelfont=scriptsize}
   \centering
    \begin{subfigure}[b]{0.45\textwidth}
        \centering
         \includegraphics[width= \textwidth]{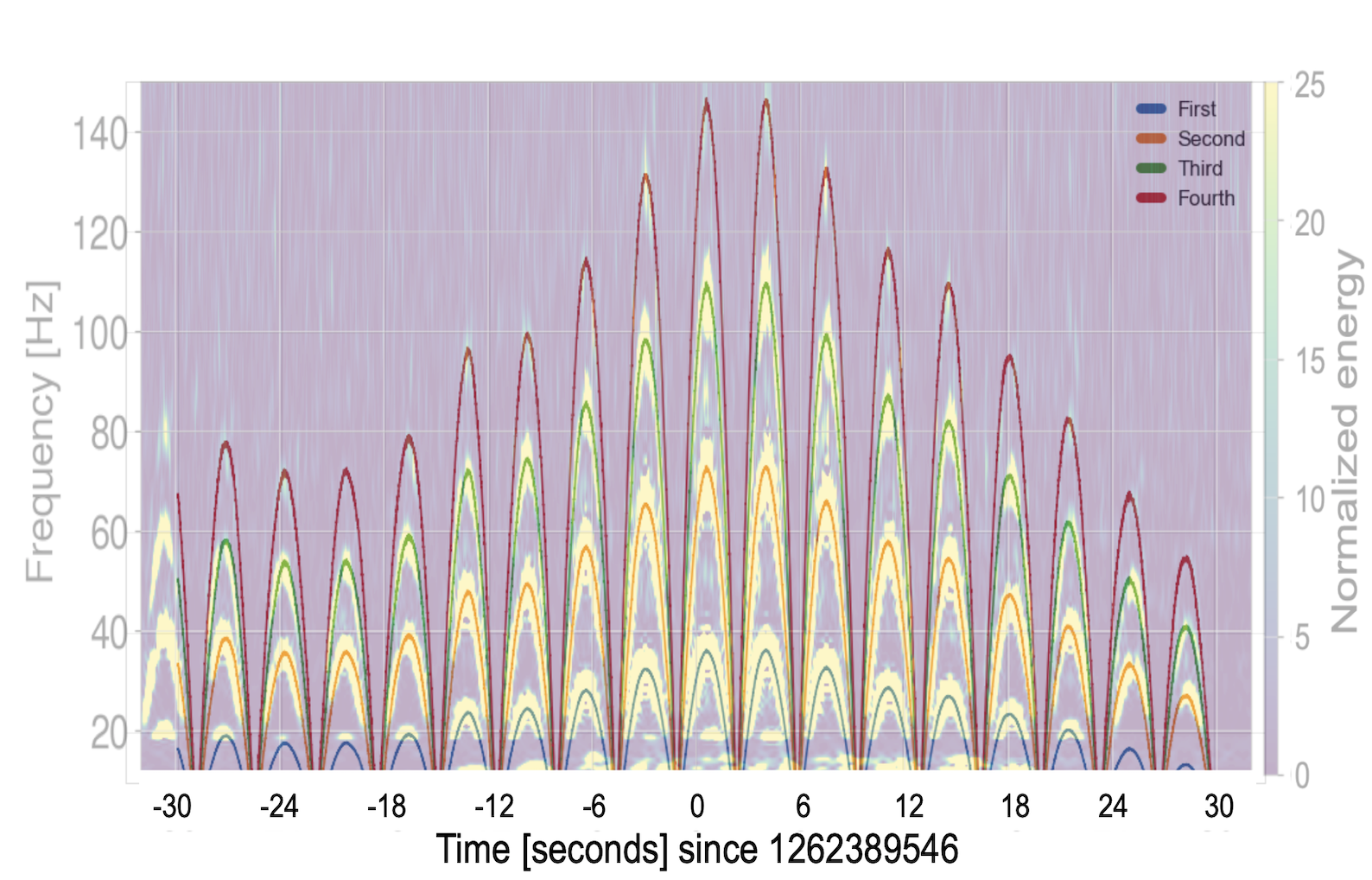}
         \caption{PUM stage fringe frequency overlaid on h(t) scattering.}
         \label{fig:fringel2}
    \end{subfigure}
    \hfill
    \begin{subfigure}[b]{0.44\textwidth}
        \centering
         \includegraphics[width =\textwidth]{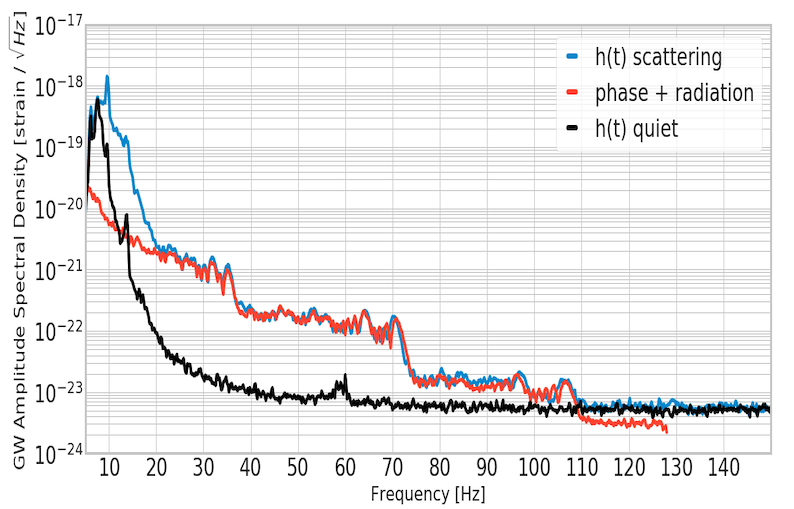}
         \caption{PUM stage amplitude spectrum density on h(t) spectrum.}
         \label{fig:spectruml2}

    \end{subfigure}
    \caption{In the left figure, we have overlaid the multiple harmonics of the fringe frequency due to the penultimate (PUM) stage motion onto scattering arches. In the right figure, we have plotted the DARM spectrum during the scattering noise shown in the left figure and the spectrum calculated from the PUM stage motion using Eq. \ref{total_noise}. DARM spectrum during a quieter time is shown as the black curve for comparison. The arches in the spectrogram on the left show up as shelves in the spectra on the right. Also, notice that the height for each successive shelf falls by a factor of 10. As we discuss in Sec. \ref{transmon_scattering}, the scattering noise in the region below 20 Hz that does not match the red curve in the right image, is due to another slow scattering coupling.}
    \label{fig:fringespectrum}
    
\end{figure}

The relative movement between the ETM and AERM, ${\delta x_{sc}}$ in Eq. \ref{eq:2}, is not directly sensed, but at low frequencies (relative to the $.45$ Hz pendulum resonance ) the motion between the ETM and AERM is similar to the motion that is sensed at the penultimate (PUM) stage of the compound pendulum. This allows us to approximate the motion between the end test mass and the reaction mass with that of PUM stage.  As shown in Fig. \ref{fig:fringel2} the fringe frequency of the PUM stage motion and its higher harmonics, calculated using Eq. \ref{eq:3}, match scattering arches in $h(t)$ spectrograms.
 
\par

On the AERM an electro-static drive (ESD) is formed by the installation of 5 gold traces \cite{esd_schematic}. Applying a voltage on the gold traces,  an electrostatic force can drive the test mass. The reaction mass is a hollow cylinder to allow the transmitted beam to pass without encountering additional optical surfaces \cite{aerm_schematic}. During operations, most of the light transmitted through the ETM goes through the reaction mass hole and onto the other side of the reaction chain as shown in Fig. \ref{fig:mirror_diag}. A small fraction of this Gaussian beam hits the gold trace ESD, on the AERM. Due to its high reflectivity, almost all of the light is back scattered towards the test mass and a fraction of that is transmitted back to the arm through the ETM. This back scattered field with an additional phase shift, given by Eq. \ref{eq:2}, interferes with the main beam in the arms and introduces phase noise in h(t).
\par
Let $E_{0}$ be the field in the arms and $E_{esd}$ is the part of this field backscattered from the end reaction mass at the point $E_{0}$ is computed. We can calculate the total field in the arms:

\begin{eqnarray}
    E_{tot} & = & E_{0} + E_{esd} \label{eq:4} \\
    E_{esd} & = & E_{0}Ae^{i{\delta}{\phi}(t)} ,\quad A = T_{end}\sqrt{f_{r}}  \label{eq:5} \\ 
    E_{tot} & = & E_{0}[1 + Ae^{i{\delta}{\phi}(t)}]  \label{eq:6}
    %E_{tot} &= E_{0}[1 + Acos({\delta}{\phi}(t)) + Asin({\delta}{\phi}(t))]
\end{eqnarray}

$T_{end}$ is the ETM  transmission ($4e^{-6}$), $f_{r}$ is the fraction of the power incident on the gold trace ESD. The calculation for $E_{esd}$ involves two transmissions through the ETM and one reflection from the ESD.

The phase noise $h_{ph}
(f)$ and the radiation noise $h_{rad}(f)$ due to this back scattered field is given by the Eq. \ref{eq:1} and Eq. \ref{rad_eqn} respectively. 
The total effective displacement power spectrum $S(f)$ can be obtained by adding the individual contributions:

\begin{equation}
S(f) = \sqrt{h_{ph}(f)^2 + h_{rad}(f)^2} \label{total_noise}
\end{equation}

Fig. \ref{fig:spectruml2} shows the total power spectrum for a scattering event on Jan 6, 2020 at LLO. The first shelf in the h(t) spectrum matches for $\sqrt{f_{r}} = 2e{-4}$, roughly consistent, within an order of magnitude with a previous estimation \cite{hiro_calc}. The coefficient for second and third shelf are $5e{-5}$ and $5e{-6}$ respectively, about an order of magnitude reduction for each higher harmonic. The amplitude of the scatter shelf for ($n+1$)th harmonic is approximately $10 \%$ of $n$th harmonic, as suggested by these reflection coefficients. This can also be observed in Fig. \ref{fig:spectruml2}.  We performed similar analysis for several scattering triggers and we found the second and third shelf coefficient magnitudes in the range of ($5e{-5}$,$6e{-5}$) and ($5e{-6}$,$6e{-6}$) respectively, while the first shelf coefficient did not vary.  
\qquad

\section{Noise mitigation using suspensions control system} \label{rzero}

Scattered light due to the large relative motion between the test mass chain and reaction chain during high ground motion has adversely affected the sensitivity of the detector. 
One way to reduce this noise coupling is by reducing the relative motion between the ETM and AERM while keeping the intended relative motion between the ETM and the input test mass (ITM) in the arm cavity. This can be achieved by sending a part of the drive from the PUM stage and feeding it to the top stage as shown in Fig. \ref{fig:end_station}. This will cause the two chains to move together and hence will reduce the relative motion between them. The reaction chain ``tracks'' the main chain and we call this RC tracking ~\cite{alog_robert}.
The reduced relative motion effectively decreases the frequency at which scattering creates shelves in $h(t)$ spectrum. 

\begin{figure}[h]
\captionsetup[subfigure]{font=scriptsize,labelfont=scriptsize}
   \centering
    \begin{subfigure}[b]{0.45\textwidth}
        \centering
         \includegraphics[width= \textwidth,height=3.9cm]{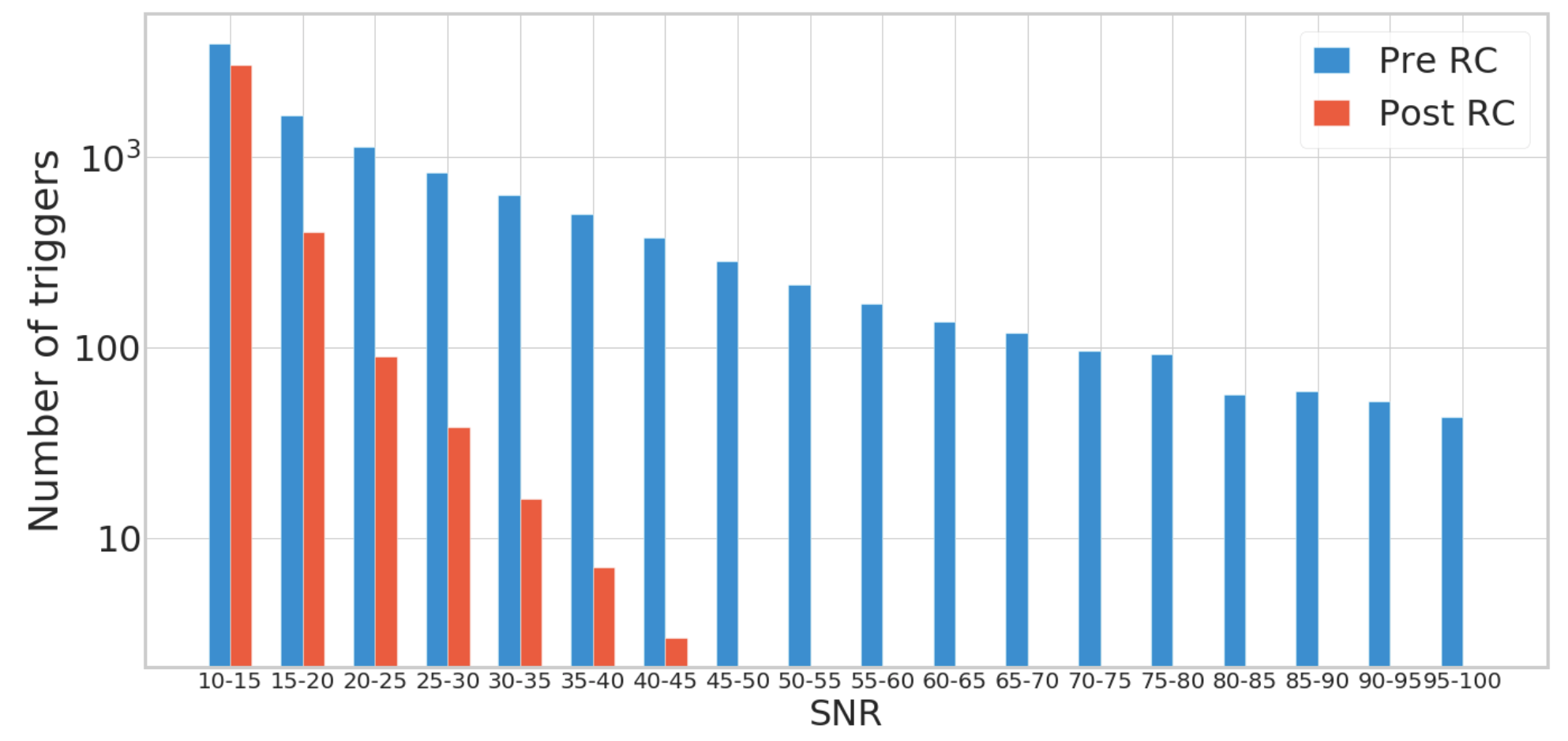}
         \caption{SNR comparison at LLO.}
         \label{fig:snr_r0LLO}
    \end{subfigure}
    \hfill
    \begin{subfigure}[b]{0.45\textwidth}
        \centering
         \includegraphics[width =\textwidth,height=3.9cm]{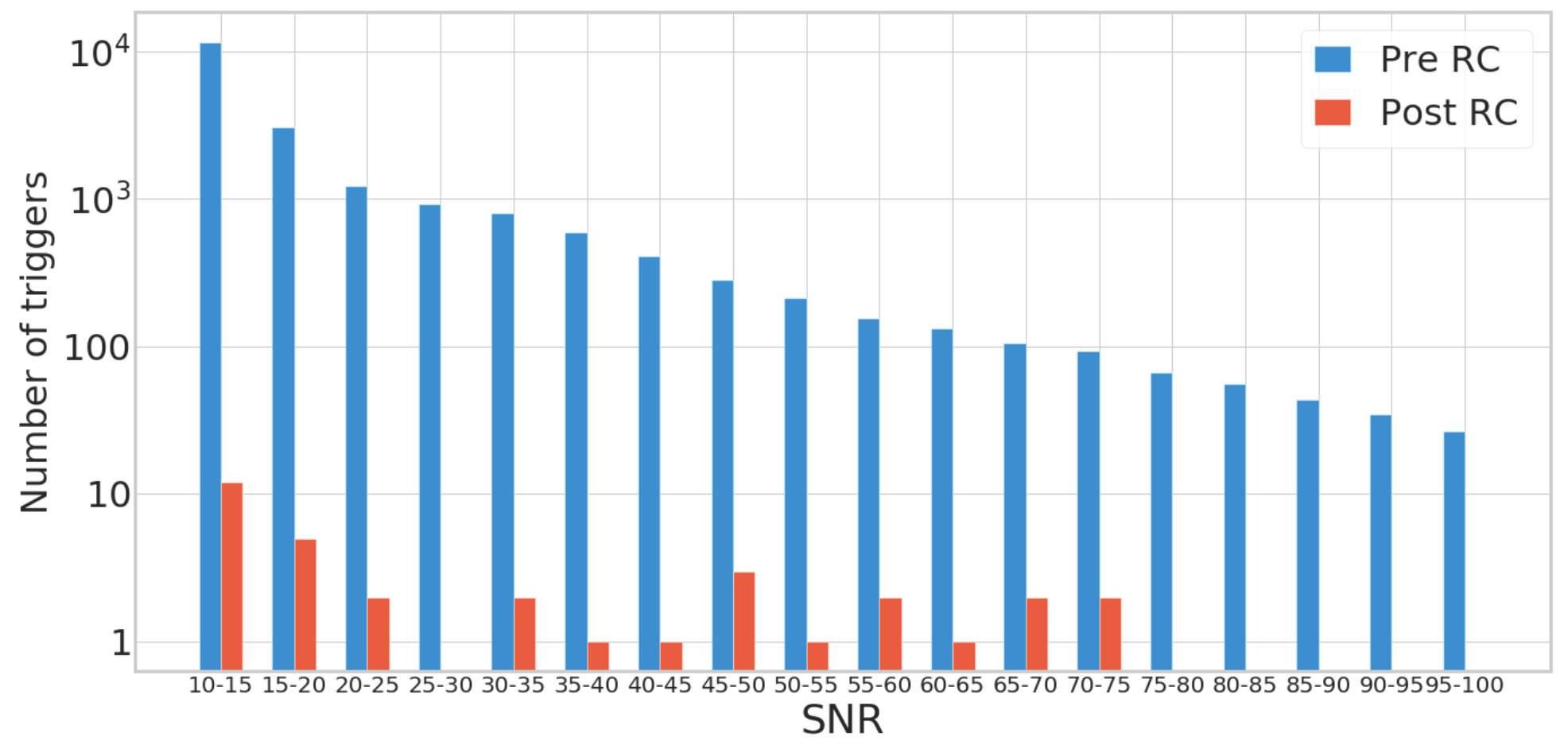}
         \caption{SNR comparison  at LHO.}
         \label{fig:snr_r0LHO}
    \end{subfigure}
    \par\bigskip
    \begin{subfigure}[b]{0.45\textwidth}
        \centering
         \includegraphics[width= \textwidth,height=3.9cm]{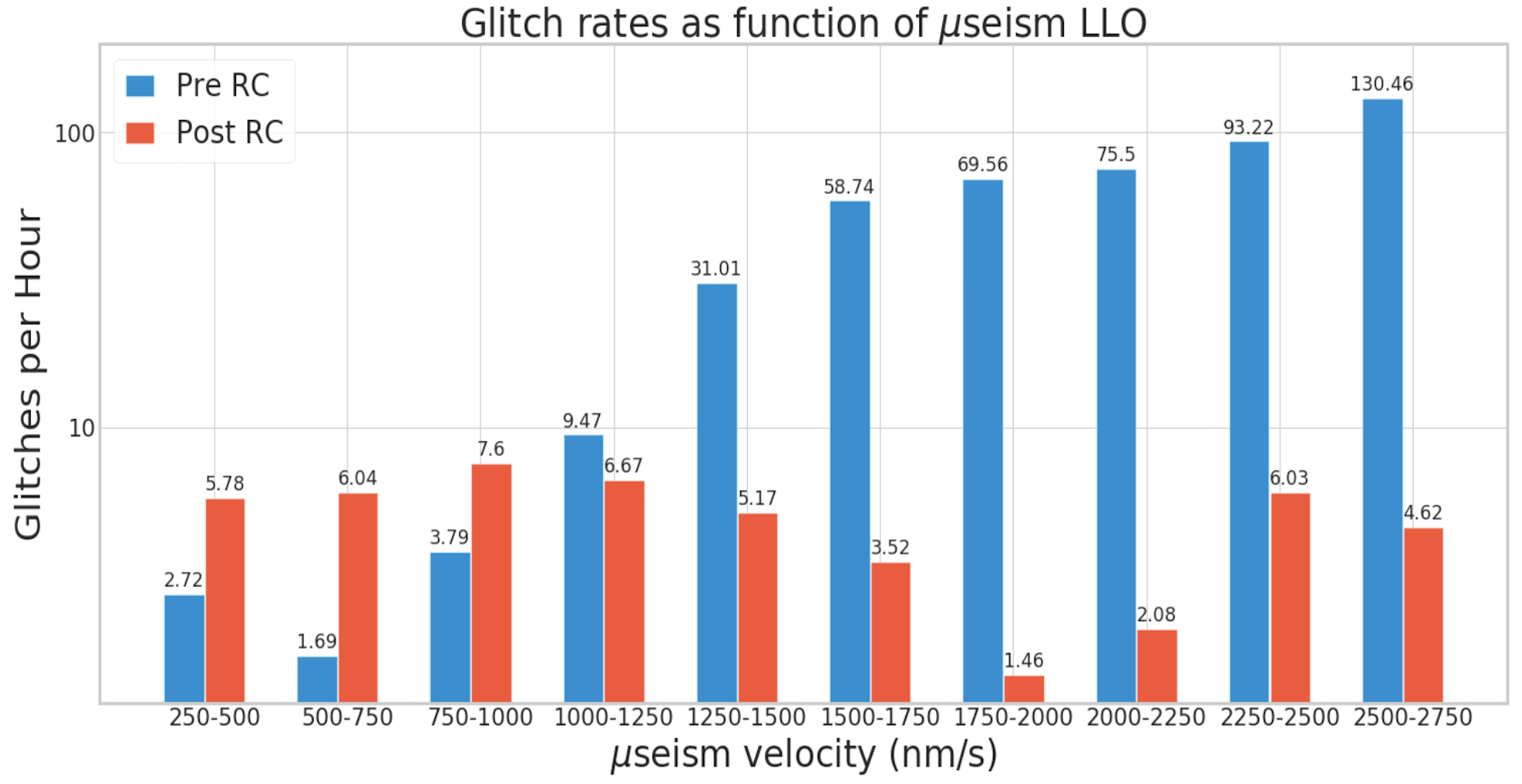}
         \caption{Glitch rate comparison at LLO.}
         
         \label{fig:rate_LLO}
    \end{subfigure}
    \hfill
    \begin{subfigure}[b]{0.45\textwidth}
        \centering
         \includegraphics[width =\textwidth,height=3.9cm]{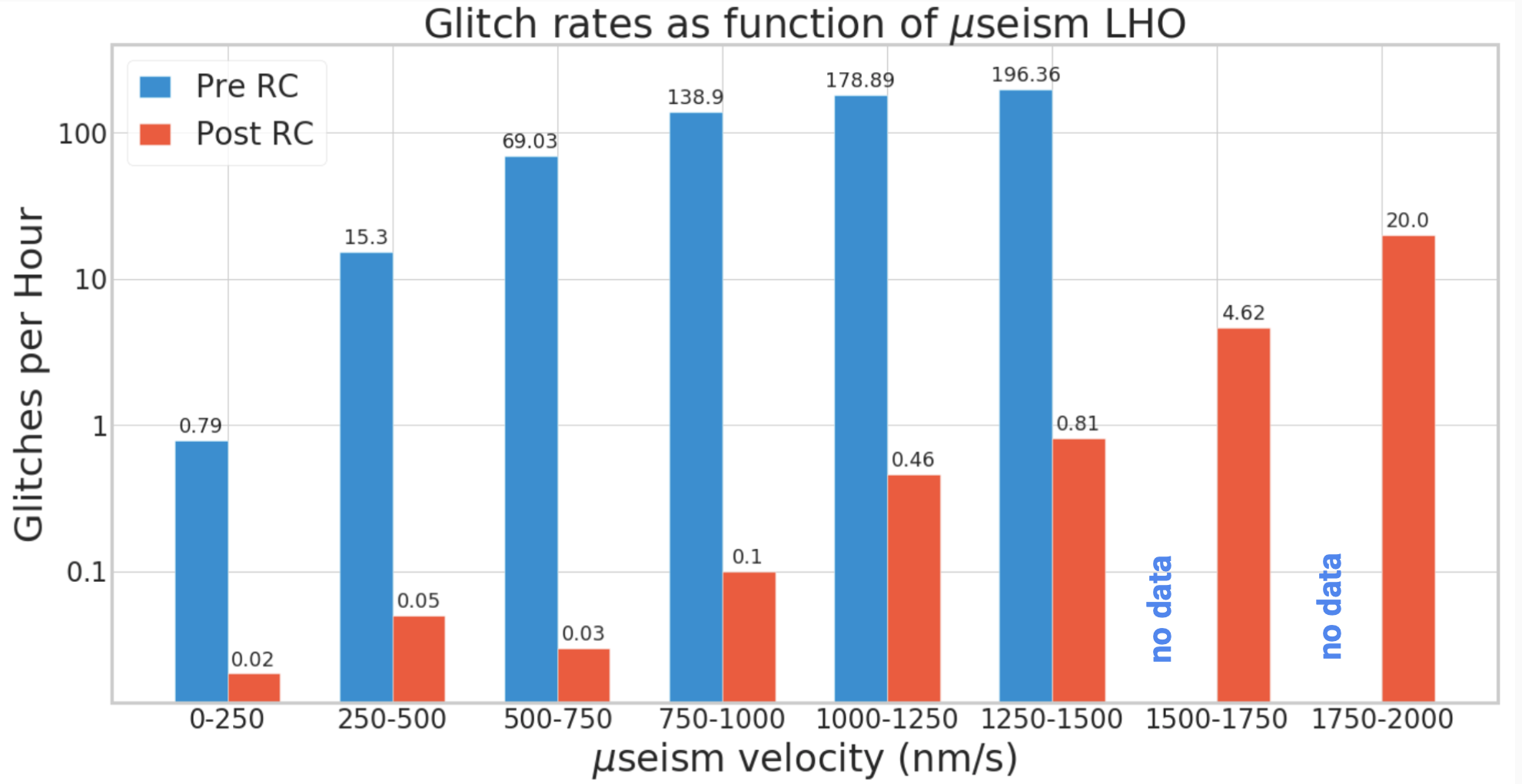}
         \caption{Glitch rate comparison at LHO.}
         \label{fig:rate_LHO}
    \end{subfigure}

    \caption{The top images compares the SNR of scattering triggers before and after RC tracking for LLO and LHO. The  SNR distribution is plotted during similar levels of before and after ground motion in the microseism band. The bottom plots compare the scattering glitch rates for different levels of microseismic ground motion before and after RC tracking. For the Pre RC tracking microseism ground motion data considered for LHO in this study, no data was found above $1500\mu{\rm m/s}$. The scattering triggers for these plots were fetched from GravitySpy with confidence above 0.9 and SNR above 10.} 
    \label{fig:postr0_scat}
    
\end{figure}

%\subsection{Reduction in  Scattering}
RC tracking was implemented on Jan 7, 2020, at LLO ~\cite{alog_anamaria}. To understand the impact of the tracking on slow scattering caused by ETM-AERM relative motion, we measured the SNR of scattering triggers and ground motion in the earthquake and microseism band between Nov 1, 2019, the start of O3b and Feb 8, 2020. We analyze triggers that are classified as scattering by GravitySpy with a confidence above 0.9 \cite{Zevin_2017}. We divided this data into Pre and Post RC, where for LLO Pre RC is from Nov 1, 2019, to Jan 6, 2020, and Post RC is from Jan 10, 2020, to Feb 8, 2020, and for LHO Pre RC is from Nov 1, 2019, to Jan 14, 2020, and Post RC is from Jan 15, 2020, to Feb 28, 2020. The analyzed data is normalized by the observing duration of Post RC considered in this study, which is $\sim 21$ days for LLO and $\sim 34$ days for LHO. Next, we considered time segments during which the ground motion in the microseismic band is similar Pre and Post RC tracking and plotted the SNR distribution of scattering triggers during these time segments. We found a clear reduction in the SNR of the scattering triggers at LLO and LHO for the Post RC scattering ~\cite{alogsid_R0}. At LLO for example, the number of triggers in the SNR bin 20-25 after RC tracking is 89, while for the same bin, before RC tracking, LLO registered 1127 scattering triggers. The SNR comparison is shown in Fig. \ref{fig:snr_r0LLO}  and Fig. \ref{fig:snr_r0LHO}.

\par
We also compared the rate of scattering triggers against the microseismic ground motion for Pre and Post RC tracking. Here again, we found that for similar levels of microseism above $1\mu{\rm m/s}$, the Post RC glitch rates are considerably lower at both the sites as shown in Fig. \ref{fig:rate_LLO}. and Fig. \ref{fig:rate_LHO}~\cite{alog_corey}. As can be seen from these figures, the ground motion in the microseismic band is usually higher at LLO than at LHO.

\section{Relative motion between test mass and transmitted light monitors} \label{transmon_scattering}
\begin{figure}[h]
\captionsetup[subfigure]{font=scriptsize,labelfont=scriptsize}
   \centering
    \begin{subfigure}[b]{0.45\textwidth}
        \centering
         \includegraphics[width= \textwidth]{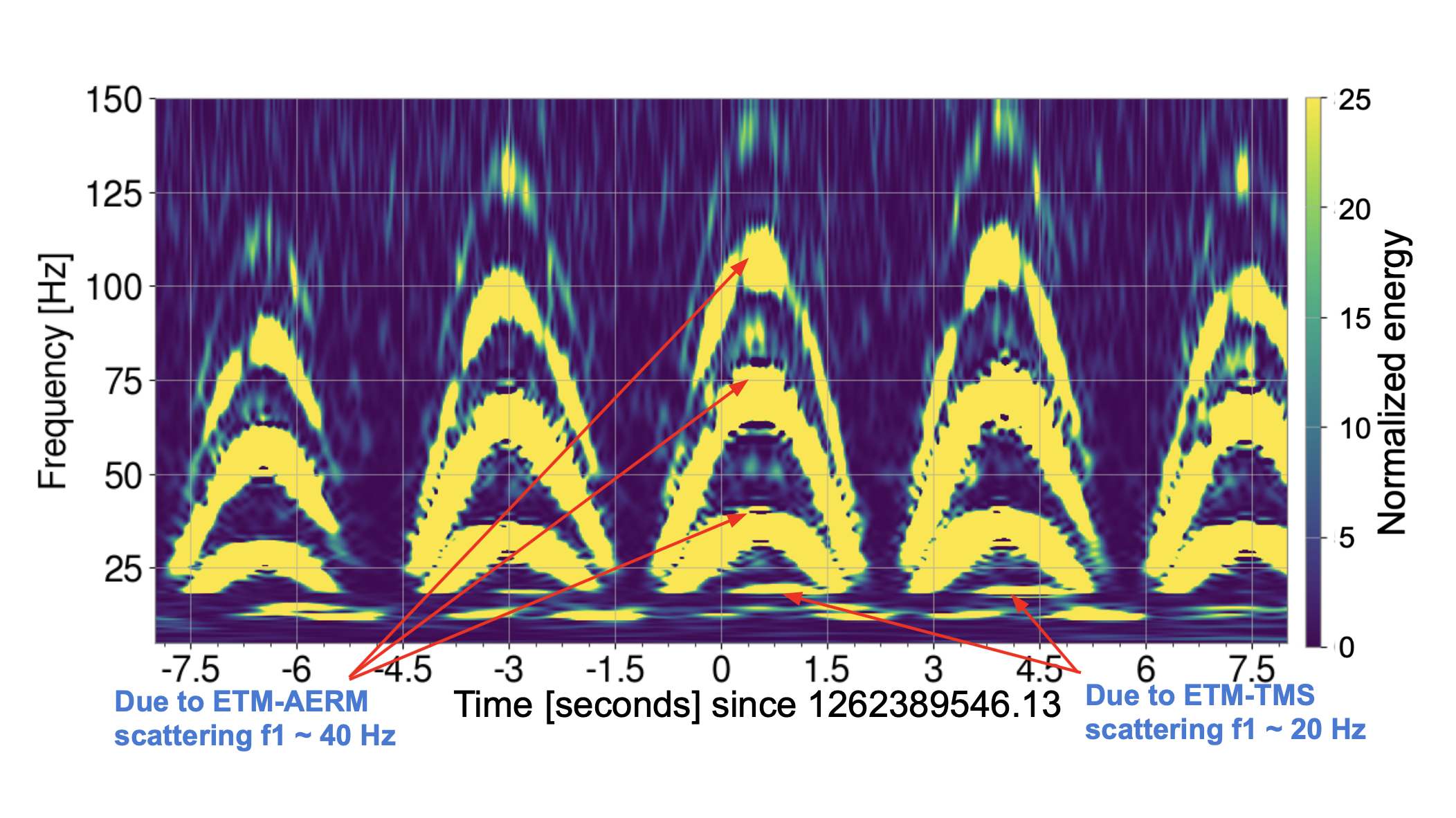}
         \caption{Scattering in h(t) before RC tracking.}
         \label{fig:1262darm}
    \end{subfigure}
    \begin{subfigure}[b]{0.45\textwidth}
        \centering
         \includegraphics[width =\textwidth]{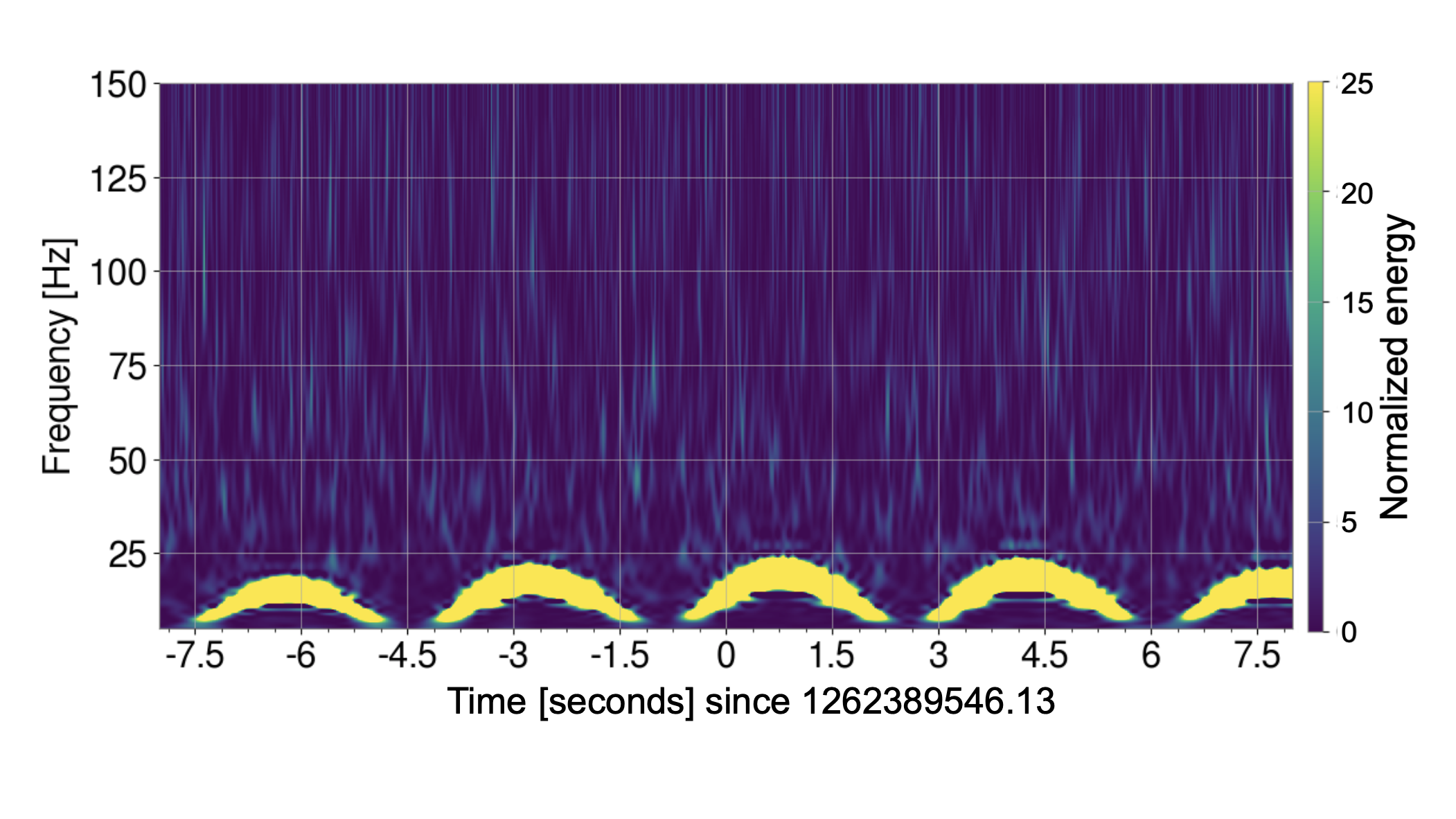}
         \caption{Scattering in X end QPD before RC tracking.}
         \label{fig:1262trans}
     \end{subfigure}
     \par\bigskip
      \begin{subfigure}[b]{0.45\textwidth}
        \centering
         \includegraphics[width= \textwidth]{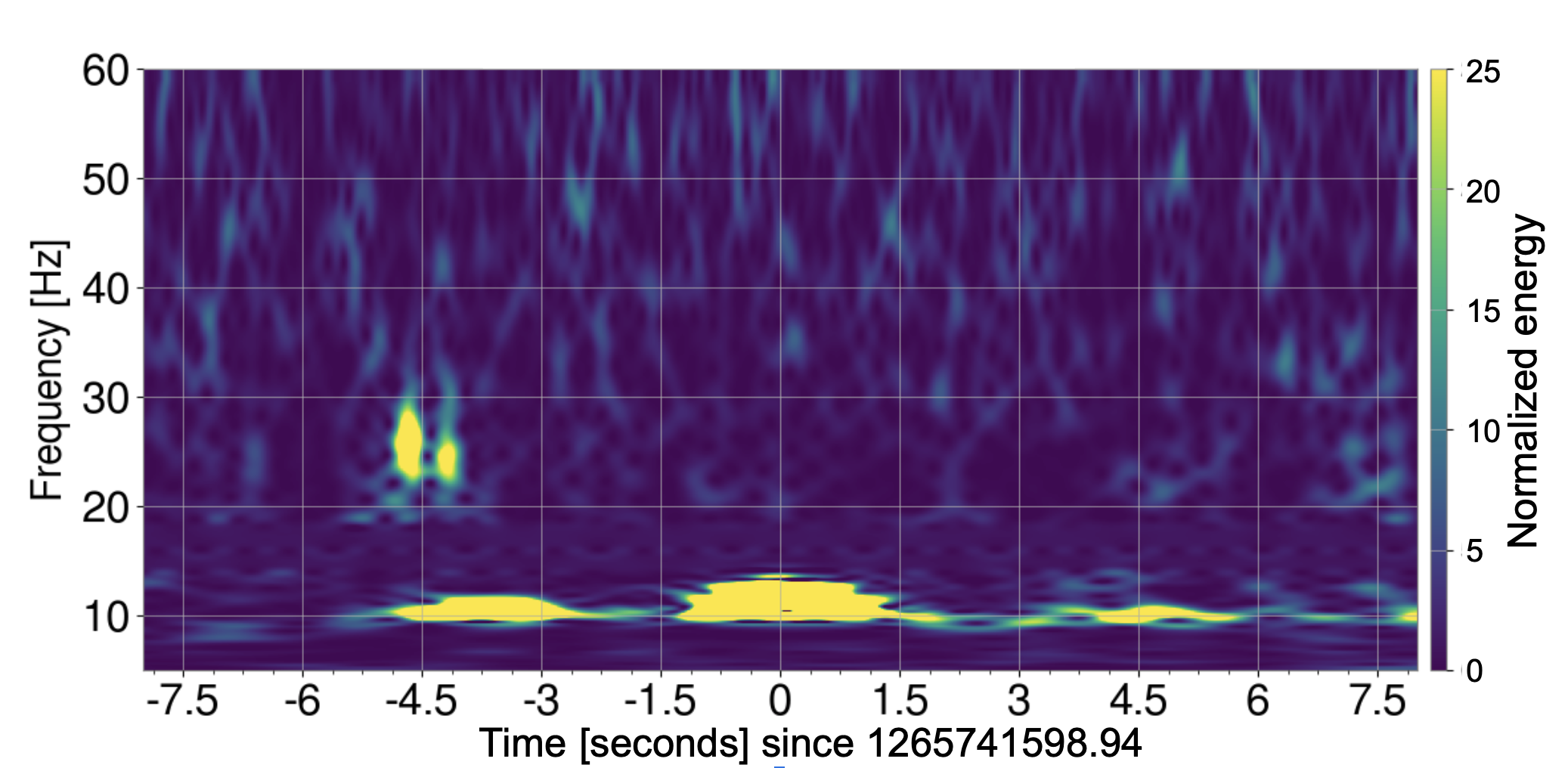}
         \caption{Scattering in h(t) after RC tracking.}
         \label{fig:1265darm}
    \end{subfigure}
    \begin{subfigure}[b]{0.45\textwidth}
        \centering
         \includegraphics[width =\textwidth]{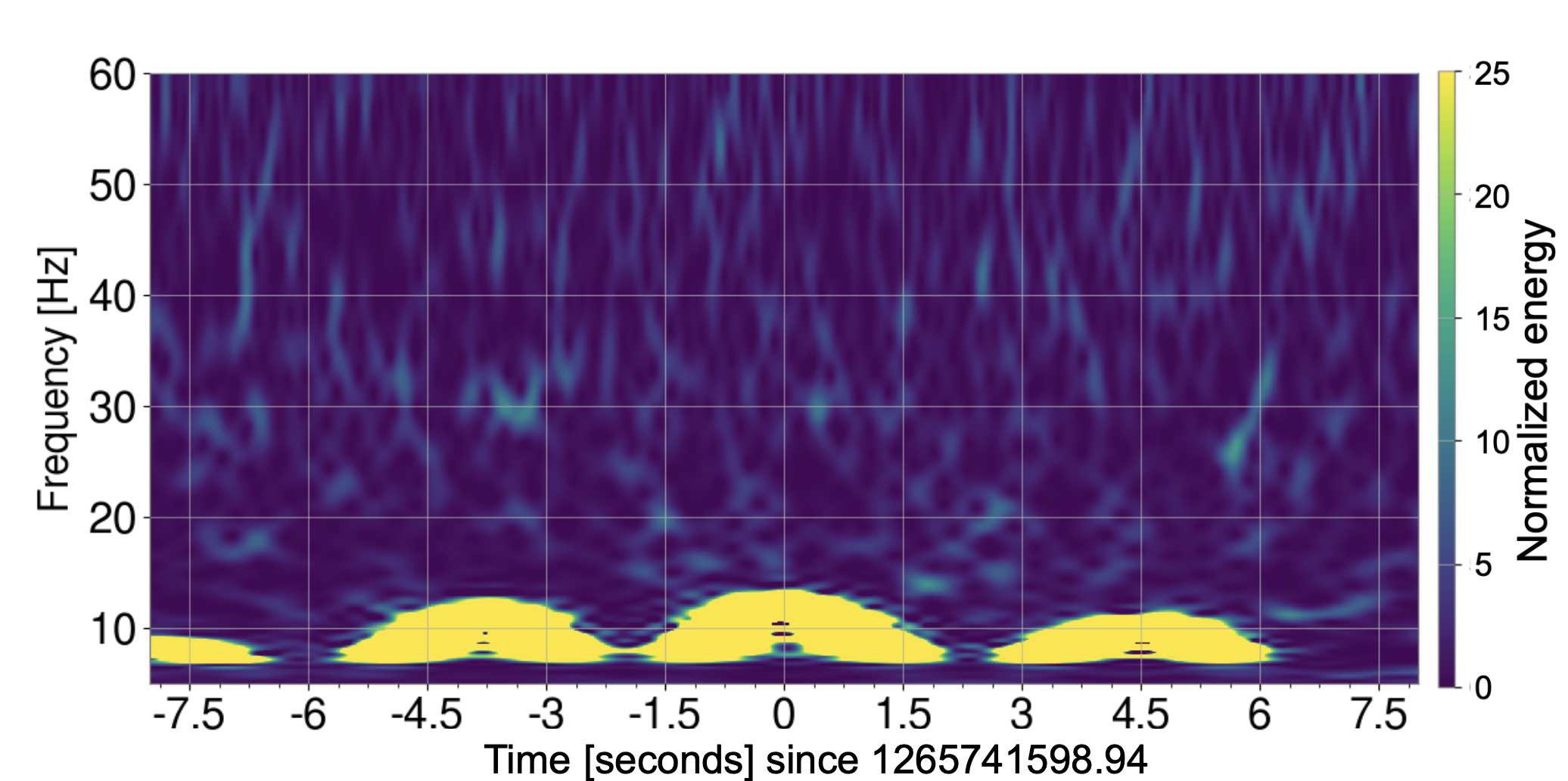}
         \caption{Scattering in X end QPD after RC tracking.}
         \label{fig:1265trans}
     \end{subfigure}

    \caption{The top left plot shows scattering arches in h(t) Q-scan during a day with very high microseism. Multiple reflections between the test mass and reaction mass generates the multiple harmonics. The first harmonic of the light scattering due to ETM-AERM relative motion is close to 40 Hz with higher harmonics present at $\sim$ 80 Hz, $\sim$ 120 Hz. Since the relative motion between the test mass and TMS is one-half of the ETM-AERM motion, the scattering arch due to ETM-TMS scattering is at $\sim$ 20 Hz. The top right plot shows this scattering arch in the transmitted light monitor. After RC tracking was implemented the noise due to ETM-AERM relative motion has reduced considerably. And thus during high ground motion post RC tracking, only the ETM-TMS noise coupling shows up as scattering arches in h(t) and the transmitted light monitor as shown in the bottom plots.}
    \label{fig:1262scat}
    
\end{figure}

As mentioned in Sec. \ref{hardware}, the control drive sent to the end test mass chain creates relative motion between the test mass and all other objects in its vicinity, such as the AERM, the TMS, vacuum chamber walls, or mechanical structures. When motion is high enough, the phase modulation from this path length difference can show up as scattering arches in both h(t) and the transmitted light monitors, labelled QPD in Fig. \ref{fig:end_station}. The motion between the main and reaction chain is twice compared to the motion between test mass and all other objects. Thus, the scattering shelf/arch due to ETM-TMS relative motion is observed at one-half the frequency of the scattering shelf/arch due to ETM-AERM relative motion. This can be seen in Fig. \ref{fig:1262darm} where the first harmonic due to ETM-AERM scattering is at 40 Hz and the scattering arch due to ETM-TMS scattering is at 20 Hz. Before RC tracking, a scattering shelf in transmitted light monitor at $f$ Hz will predict scattering shelves in h(t) at $f$ Hz due to ETM-TMS coupling and at 2$f$, 4$f$, 6$f$ and so on due to ETM-AERM coupling. Following RC tracking, a shelf in transmitted light monitor at $f$ Hz only corresponds to a shelf in h(t) at the same frequency.
\par
 
\begin{figure}[h]
    \centering
    \includegraphics[width=\textwidth]{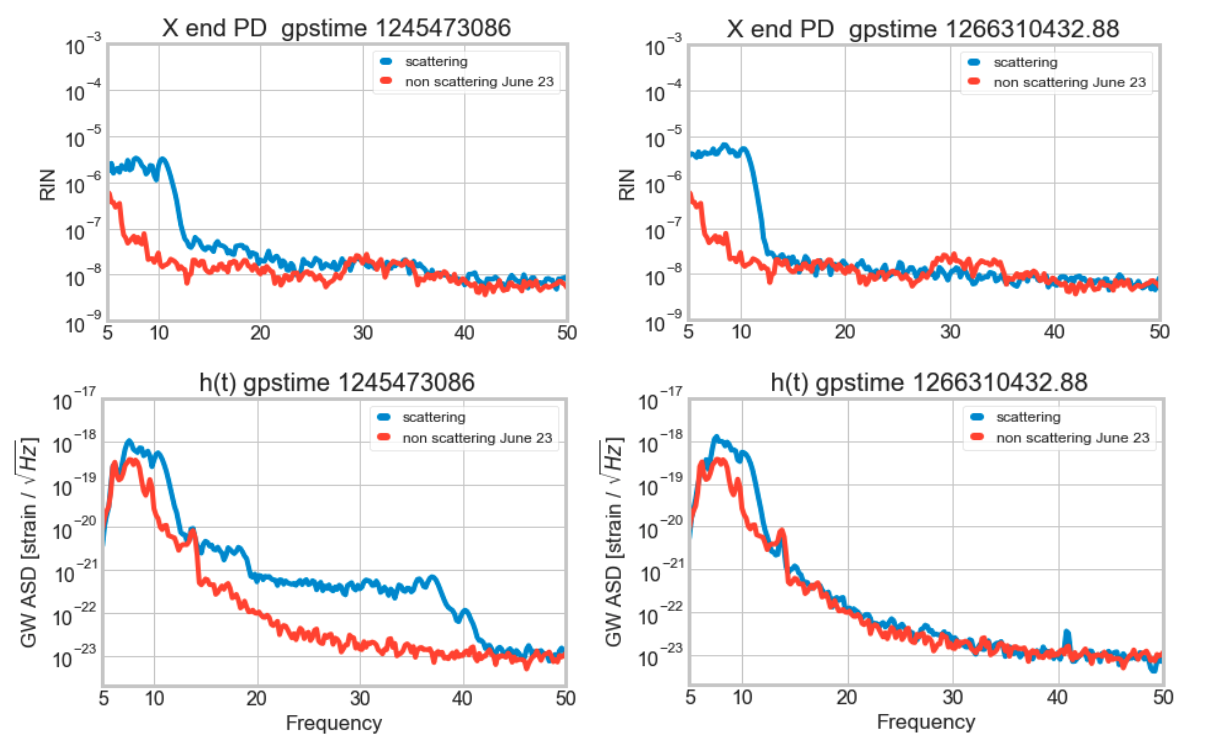}
    \caption{The left image shows Relative Intensity Noise (RIN) of  X end transmitted light monitor and h(t) before RC tracking was implemented. The scattering shelf at 10 Hz in h(t) correlates with ETM-TMS noise coupling, while the higher frequency shelves at $\sim$ 20 Hz and $\sim$ 40 Hz are due to ETM-AERM scattering. The image on the right is for a Post RC tracking scattering event and higher frequency shelves are absent in h(t) due to reduced relative motion between the main chain and the reaction chain.}
    \label{fig:etm-transmon}
\end{figure}

\par
Fig. \ref{fig:1265darm} shows a slow scattering arch in the time-frequency representation, after the RC tracking was implemented. The lack of multiple arches suggests that the scattering path does not involve multiple traversals between the test mass and the scatterer. The TMS-ETM scatter mechanism was confirmed experimentally. Low frequency motion was injected at the Y end seismic isolation table, forcing the DARM loop to respond by inducing large motion at the X end test mass and creating a scatter shelf in h(t). The TMS was then fed the same motion, reducing the relative motion in between it and the optic and the h(t) scatter shelf disappeared ~\cite{alogae_tmsx,alogsid_tmsx,Heijningen:2018evm}.

With RC tracking the higher frequency scattering shelves due to ETM-AERM coupling have gone away. Fig. \ref{fig:etm-transmon} compares the scattering shelves in h(t) and X end transmitted light monitor for scattering events before and after RC tracking. The DARM control signal is sent to one test mass and this results in large relative motion between the test mass and its surroundings. One remedy to reduce the ETM-TMS relative motion is to split and apply this control drive at all four test masses forming the LIGO arm cavities. This will reduce the relative motion by a factor of 4. Further reduction can be employed by making the TMS follow the test mass chain like in the test described above and we intend to implement this for the next observing run.

\par 

\vspace{0.5cm}
 
\section{Using transmitted light monitors to identify scattering}\label{transmonwitness}
GravitySpy is an image recognition tool that uses machine learning to classify the variety of omicron triggers that show up as transient noise in the strain data.  It is a citizen science project and volunteers help to generate the training dataset by assigning one of the several glitch classes to the spectrogram images. The algorithm assigns each image a glitch class and a confidence score which represents the probability that the image belongs to that specific glitch class.
We can identify times of transients due to scattered light by looking at the output of GravitySpy \cite{Zevin_2017,gspymachine}.

\subsection{GWDetChar-Scattering}
\begin{figure}[h]
\captionsetup[subfigure]{font=scriptsize,labelfont=scriptsize}
   \centering
    \begin{subfigure}[b]{0.45\textwidth}
        \centering
         \includegraphics[width= 7cm]{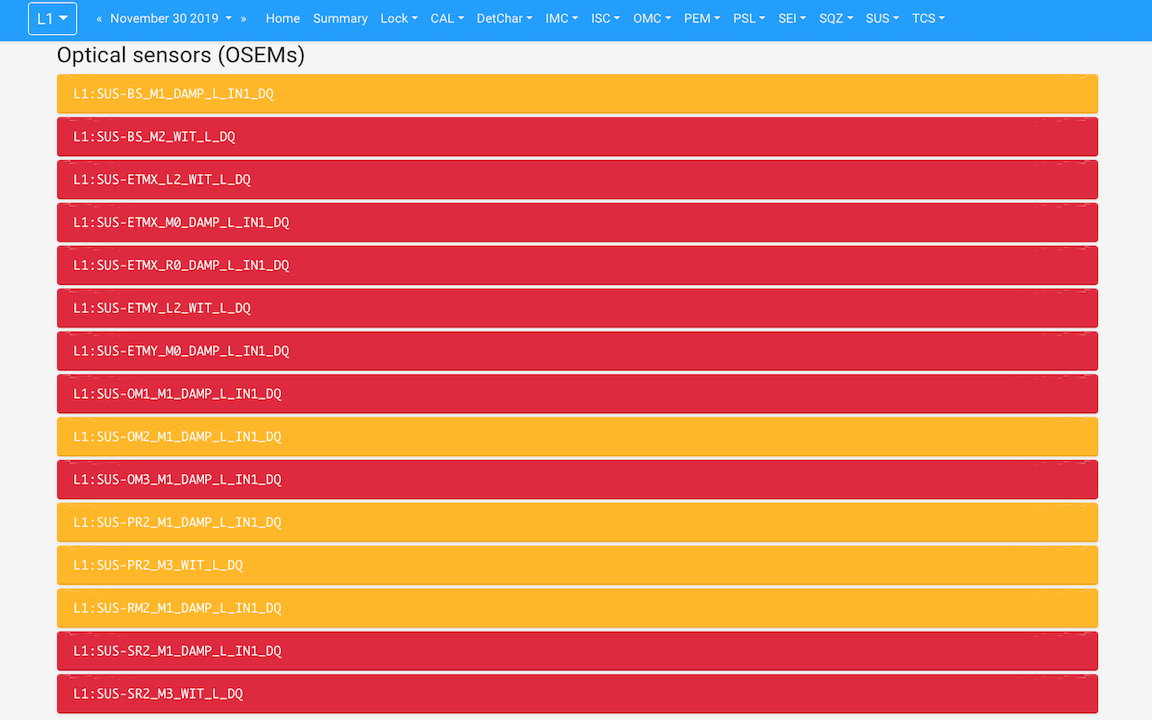}
         \caption{Optical sensors on scattering summary page.}
         \label{fig:summary_scattering}
    \end{subfigure}
    \hfill
    \begin{subfigure}[b]{0.5\textwidth}
        \centering
         \includegraphics[width =7cm]{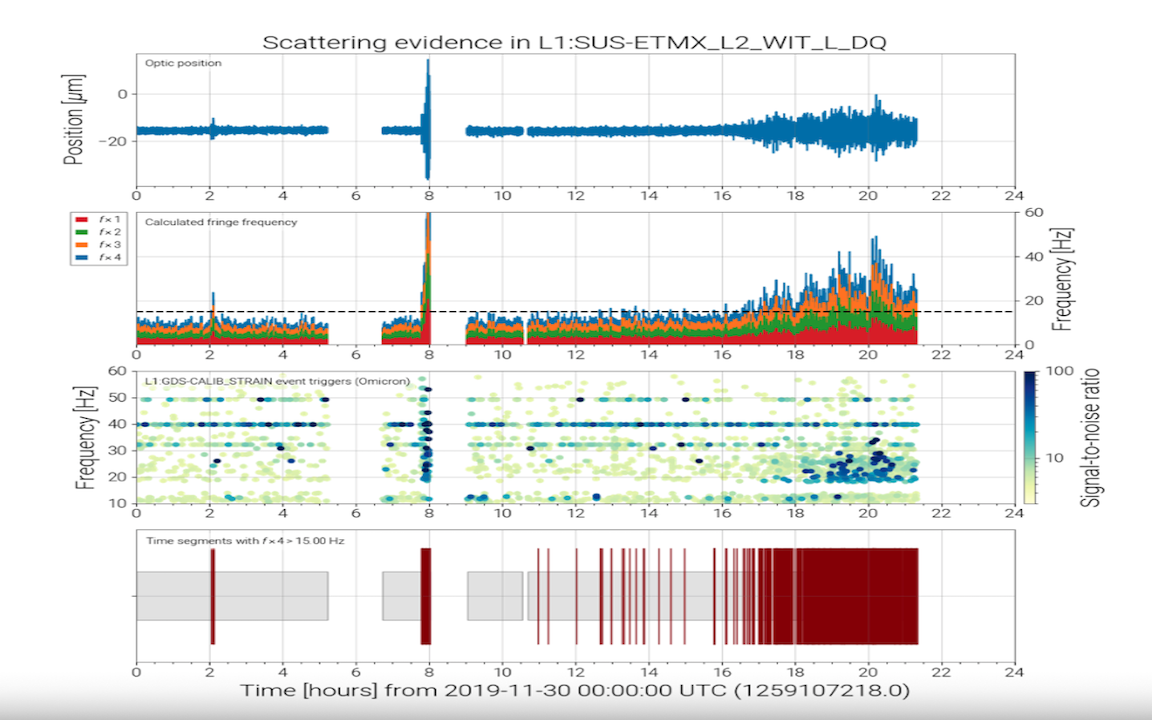}
         \caption{ Scattering evidence in OSEMs.}
         \label{fig:optic_motion}

    \end{subfigure}
    \caption{The left image shows the optical sensors processed by \textbf{gwdetchar-scattering} for evidence of scattering. They are color coded with red for strong evidence of scattering and yellow for weak. The topmost plot on the right image shows the position measurement by the OSEM on the PUM stage of the reaction chain, followed by the fringe frequency calculated using \ref{eq:3}. The third plot shows h(t) Omicron glitches with the glitch rate increasing with the increase in fringe frequency motion. The final plot shows the time segments during which the fringe frequency motion is above 15 Hz. The h(t) triggers that lie within these scattering segments are written to a file as potential scattering.}
    %During notably high micro-seismic motion, there are several more slow scattering arches, that causes the third bump in the duration plot.}
    \label{fig:summary_optic}
    
\end{figure}

Another way of identifying the potential scattering triggers is by monitoring the motion of OSEMs and then correlating it with the presence of the same triggers in the gravitational wave strain channel. This is accomplished by using an algorithm called - \textbf{gwdetchar-scattering}~\cite{gwdetchar}.  Throughout the detector, OSEMs are used to capture the motion of any components, light can get scattered from. The timeseries data from these sensors is used to find potential site of scattering in the following way. The position measurements taken by the OSEMs is first converted velocity using a savgol filter \cite{2020SciPy} and then to fringe frequency using the following equation \ref{eq:3}.
The \textbf{gwdetchar-scattering} script then creates segments of the form (startime, endtime) during which the fringe frequency motion in optic crosses a certain frequency threshold. The algorithm then looks for time coincidences between these segments and omicron triggers in h(t) in 10 Hz - 60 Hz frequency band. Efficiency is defined as the percentage of triggers in h(t) channel that falls within these segments while the deadtime is the duration of all the segments for an optic as a percentage of total observing time. An optic is considered to be ``strong'' witness if the ratio of efficiency over deadtime is greater than 2 and ``weak'' if it is less than 2. The script then prepares a webpage, as shown in Fig \ref{fig:summary_optic}, showing the movement of all the optics, the scattering segments of each optic and the information of h(t) triggers captured by these segments \cite{ligo_summary}.  

Another method to identify scattering culprits is to employ an adaptive algorithm based on time varying Empirical Mode Decomposition (tvf EMD) used at Virgo \cite{Longo_2020}. This method which utilizes the non-linear nature of light scattering, finds correlation between the Instantaneous Amplitude (IA) of primary channel and time derivative of potential scatterer's position. A method based on Hilbert Huang transform has also been developed to catch scattering surfaces \cite{Valdes_2017}. The tvf EMD and the Hilbert Huang  methods are based on quantifying time series correlation between the gravitational strain data and potential scattering surface. The \textbf{gwdetchar-scattering} finds time coincidence between moving optical surfaces and trigger data processed by another pipeline (Omicron).

GravitySpy, even though does not provide any information with regards to where the scatterer might be located, identifies a larger subset of scattering triggers compared to that identified by motion in OSEMs. On the other hand, optics motion can be a more direct method of locating the source of scattering noise since it can identify which mirror is moving with the velocity required. It thus makes sense to see if we can make \textbf{gwdetchar-scattering} more ``efficient'' by adding better scattering witnesses to the algorithm. In this section, we explore such a witness that can be used to identify the scattering noise. 

\begin{figure}[h]
  \centering
         \includegraphics[width = 14cm,height=12cm]{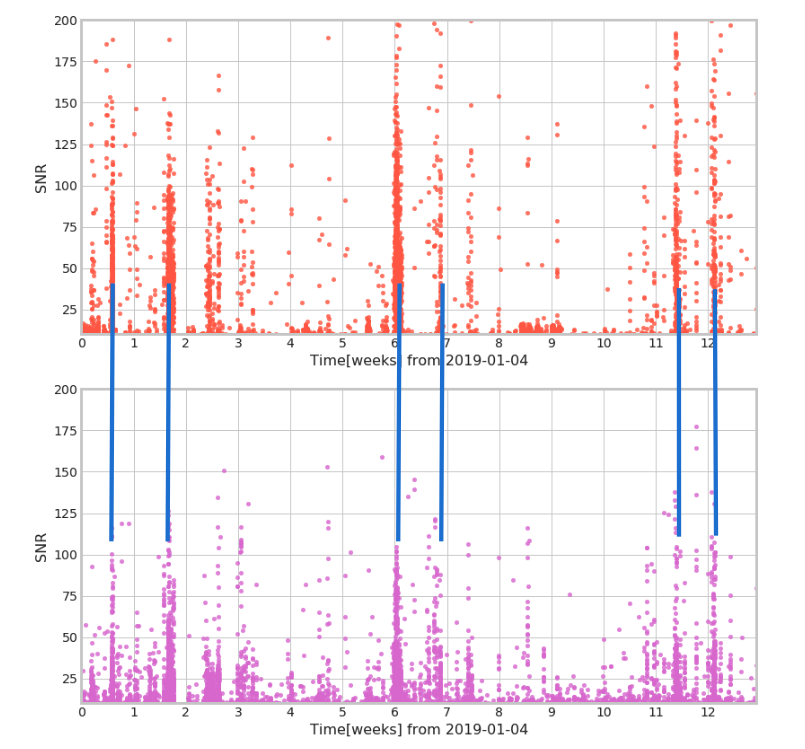}
         \caption{Time coincidence between X end transmitted light monitor and slow scattering in h(t) channel during the first three months of O3a at LLO. The top plot shows slow scattering triggers identified by GravitySpy above a confidence of 0.95 in h(t) and the bottom plot shows the omicron triggers in X end transmitted light monitor during the same time period. The vertical lines show the correlation between the scattering in h(t) channel and noise in the transmitted light monitor channel.}
    \label{fig:transetmx}
\end{figure}

In Sec. \ref{transmon_scattering}, we showed that the transmitted light monitors serve as a witness of slow scattering noise in h(t). Fig. \ref{fig:transetmx} shows a time correlation between the slow scattering triggers in h(t) as identified by GravitySpy and the noise in the transmitted light monitor below 20 Hz. Due to the presence of this temporal coincidence of triggers, the noise in this auxiliary channel can be used to identify the slow scattering noise in h(t). 
%Along with OSEM’s, these channels can be added to the list of scattering witnesses in the algorithm - \textbf{gwdetchar-scattering}.
These channels can be added to the list of optics, over which the algorithm \textbf{gwdetchar-scattering} iterates to find scattering in h(t). This change represents an update on \textbf{gwdetchar-scattering} rather than a new algorithm to find scattering noise. 

Apart from slow scattering, ``Extremely Loud triggers", which is another class of triggers characterized by typically high SNR, also makes its way to the transmitted light monitors. The source of these loud triggers is not well understood but it is not believed to be related to scattered light, so we need to remove these triggers from our analysis. To differentiate the presence of slow scattering noise in h(t) from loud triggers, we can look at the frequency content of the coincident noise in the transmitted light monitors. The scattering noise in these witness channels appears in the range  4 - 10 Hz while the triggers coincident with loud glitches in h(t) typically appear with higher peak frequency. 

\subsection{Band-limited RMS}
\begin{figure}[h]
    \centering
    \includegraphics[width=10cm]{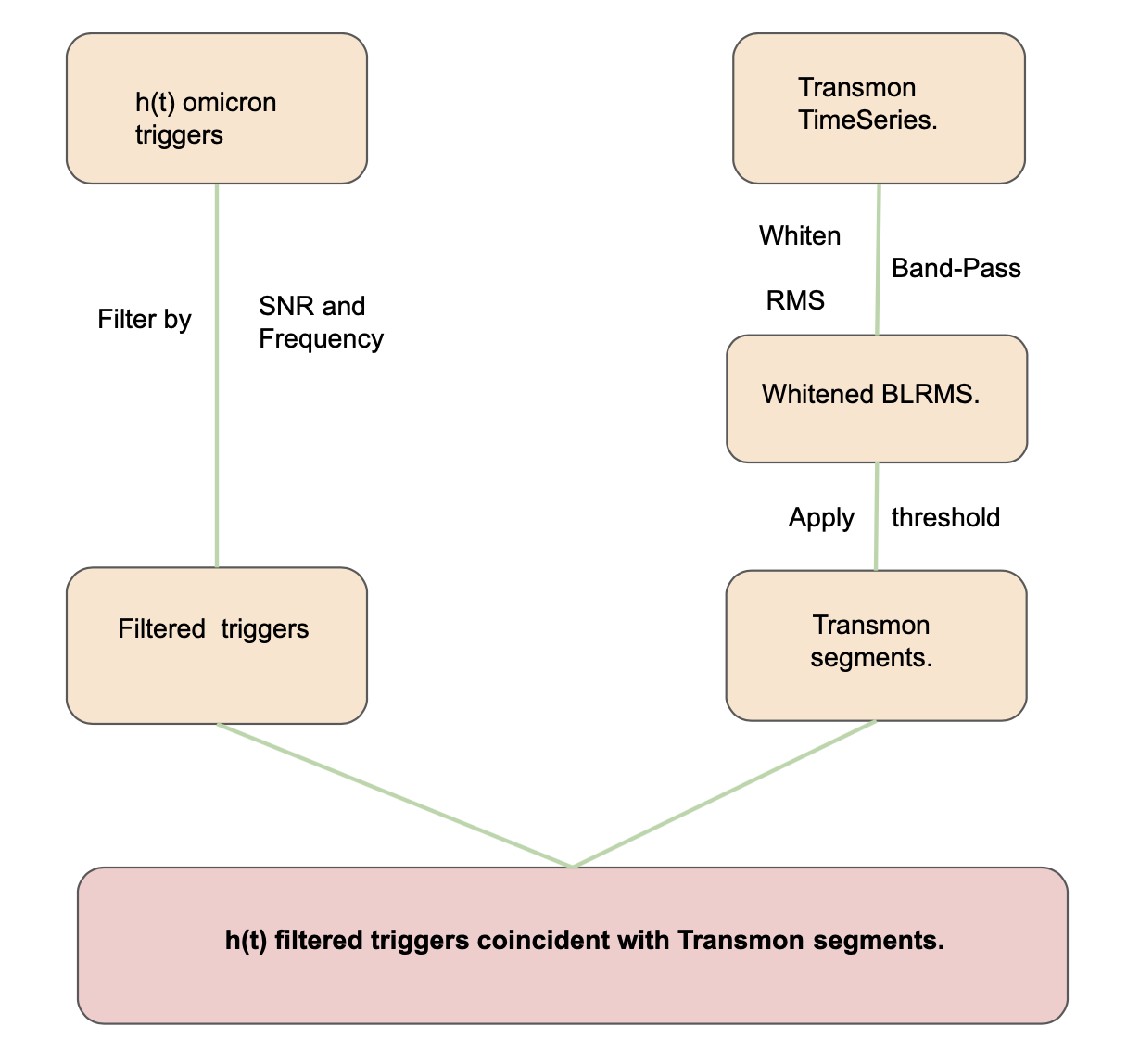}
    \caption{Flowchart of the process to capture scattering in h(t) by using segments generated from whitened transmon BLRMS. Transmon or transmitted light monitor's time series is first whitened to normalize the power in the frequency bins. This is followed by band-passing the data between 4 and 10 Hz. Since the scattering arches show up in this frequency range in the transmon, band-passing will remove any noise outside this range. After taking the root-mean-square, any values of this whitened and band-passed time-series above a given threshold are converted to segments. We then look for coincidence between these BLRMS segments and h(t) omicron triggers filtered by SNR and frequency.}
    \label{fig:flow}
\end{figure}

To capture the scattering triggers in h(t), we use whitened band-limited root mean square (RMS) segments constructed from the raw time-series of the transmitted light monitors, in the frequency band of interest. A scattering trigger in h(t) shows up as a spike in these band-limited RMS (BLRMS). By choosing a suitable threshold, we can create BLRMS segments and then use time coincidence with the Omicron triggers in h(t) to identify the scattering triggers. Any h(t) triggers that coincide with these band-limited segments are then written to a file as potential scattering.  This process is shown in Fig. \ref{fig:flow}. Before finding time coincidence between the h(t) triggers and BLRMS segments, we filter the triggers by SNR (between 15 and 200) and frequency (between 10 Hz and 60 Hz), thus excluding the loud triggers that can pollute the algorithm. The transmitted light monitors witness loud slow scattering as $95$ \% of the light transmitted through the ETM is dumped before it reaches the TMS, hence we use a lower SNR limit of 15.

\quad

%\subsection{Results}

%\par
We performed the analysis for O3a, from April 1, 2019, to Sep 30, 2019. The BLRMS segments identified 3864 h(t) triggers as scattering at LLO while GravitySpy found 3663 scattering triggers above confidence of 0.8.  Three-fourth of the GravitySpy scattering match with scattering triggers caught by BLRMS segments. 71$\%$ of the 3864 triggers caught by BLRMS segments match with GravitySpy output. This suggests that 29$\%$ of 3864 or approximately 1120 triggers are false positive with respect to GravitySpy. 
The time-frequency spectrograms of 57 randomly chosen trigger times from these 1120 triggers showed that as many as 40 of these were scattering triggers, but they were not labelled as scattering by GravitySpy above confidence of 0.8.

\begin{figure}[h]
   \centering
    \begin{subfigure}[b]{0.45\textwidth}
        \centering
         \includegraphics[width= \textwidth]{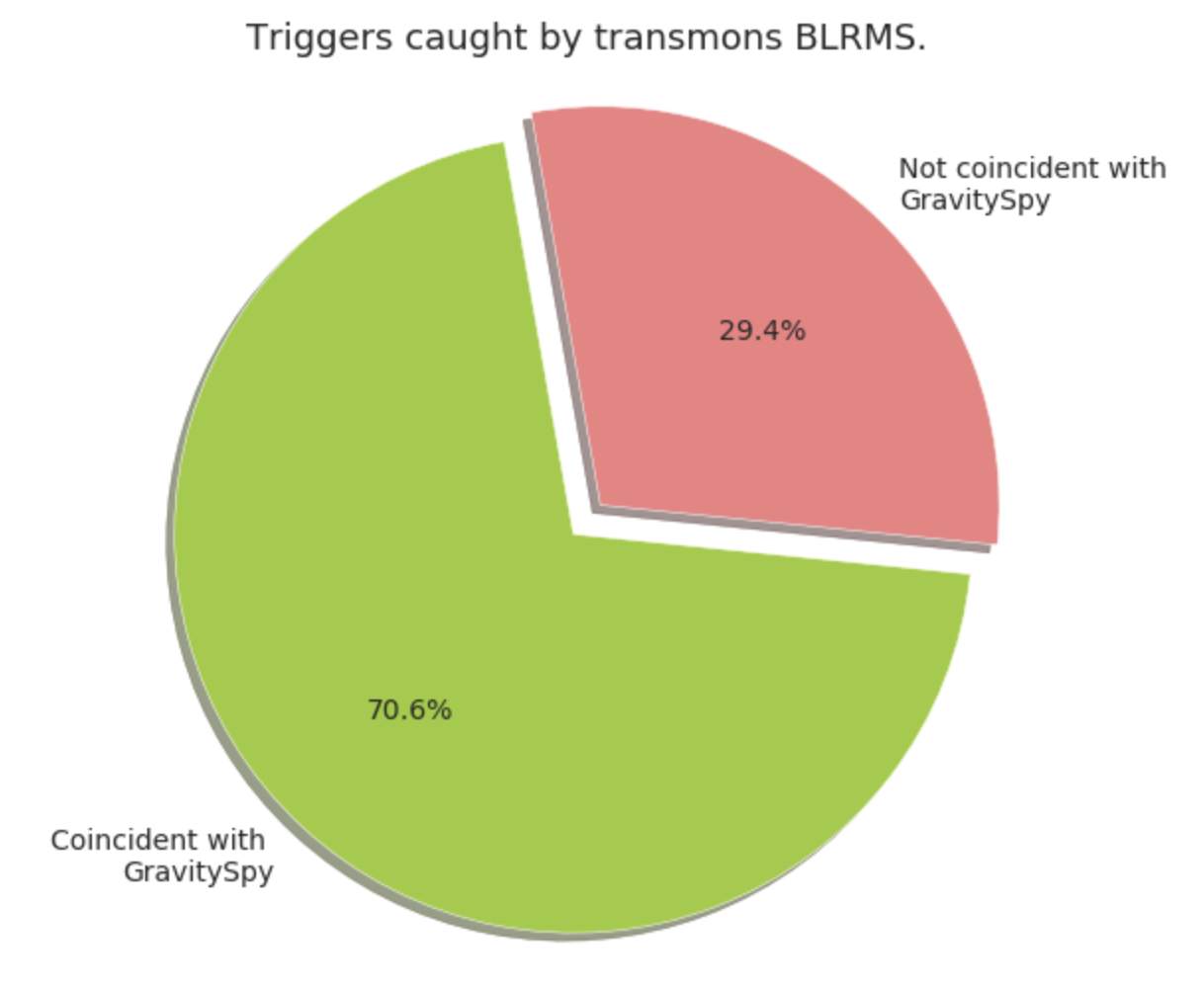}
         \caption{}
         \label{fig:transpie}
    \end{subfigure}
    \hfill
    \begin{subfigure}[b]{0.44\textwidth}
        \centering
         \includegraphics[width =\textwidth]{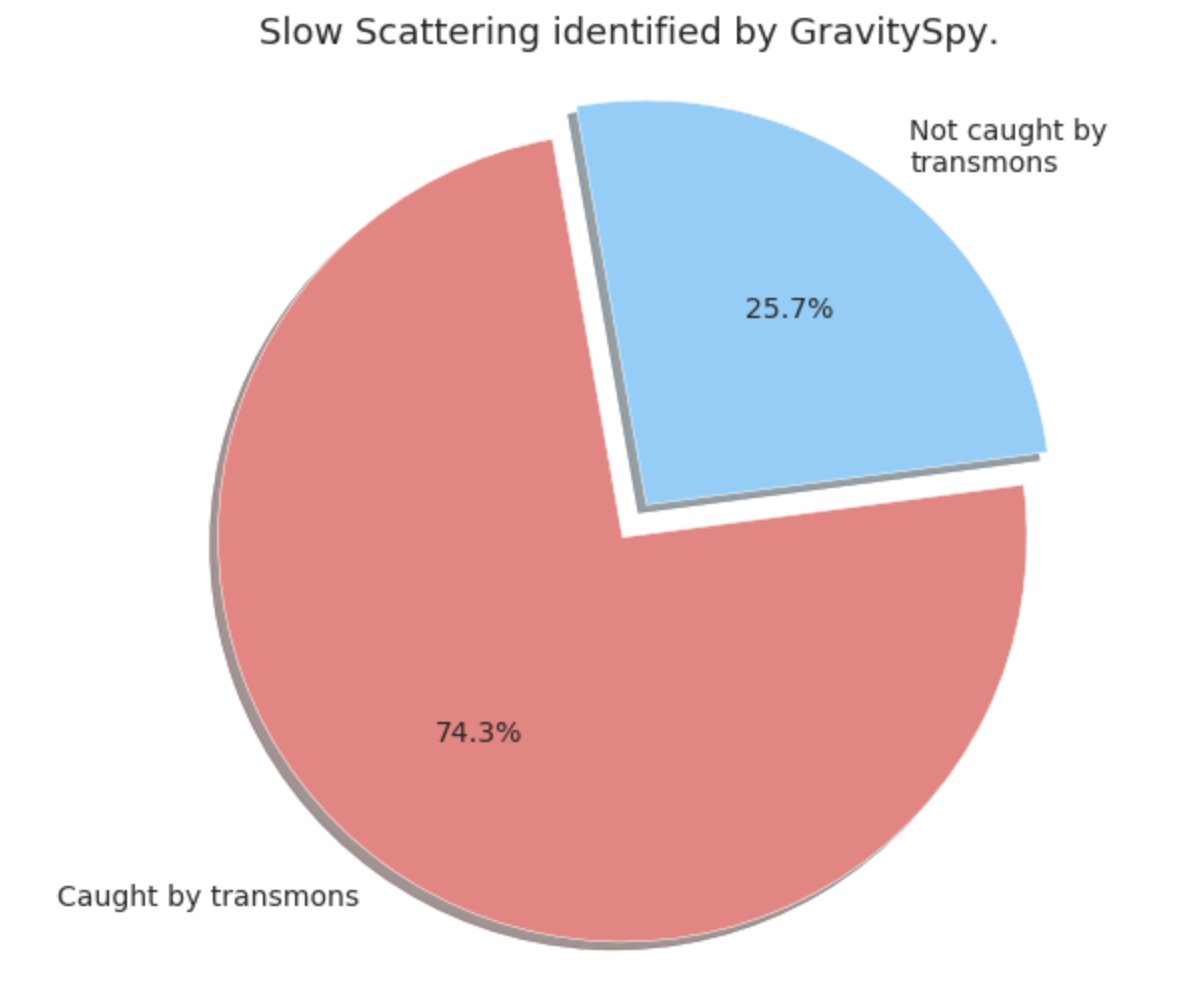}
         \caption{}
         \label{fig:gspypie}

    \end{subfigure}
    \caption{For O3a, the first pie chart shows that most of the triggers vetoed by  BLRMS segments are identified as scattering by GravitySpy above a confidence of 0.8. The spectrograms of a subset of false positives show most of them are slow scattering but were labeled with a confidence lower than 0.8 by GravitySpy. The second pie chart shows that BLRMS segments caught close to 75 \% of the slow scattering that GravitySpy identified above a confidence of 0.8.}
    \label{fig:gspytrans}
    
\end{figure}
\par

\begin{figure}[h]
    \centering
    \includegraphics[width=10cm]{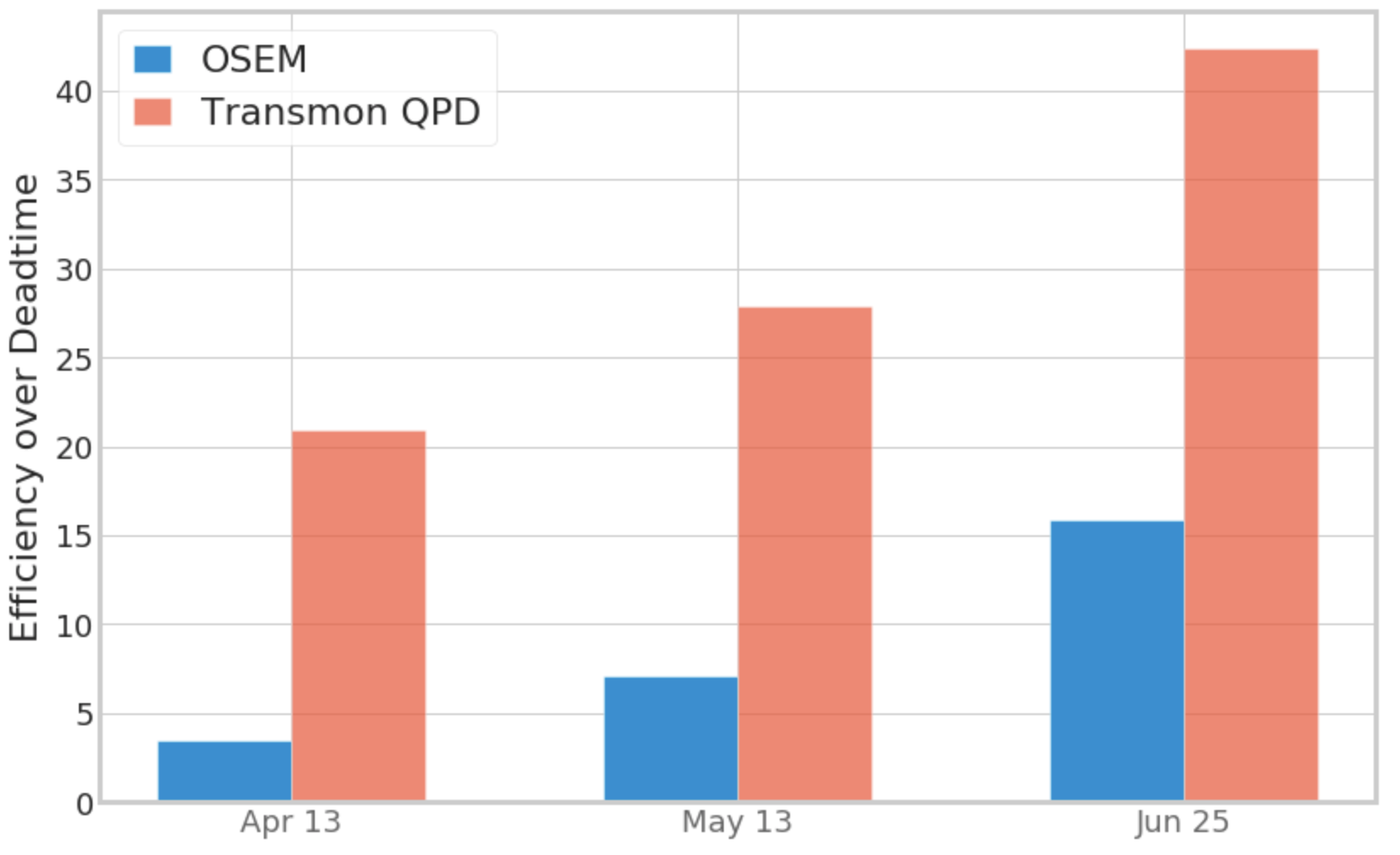}
    \caption{Comparing efficiency over deadtime between OSEMs and BLRMS segments. This shows that for slow scattering that occurred on Apr 13, May 13 and June 25, 2019, transmmited light monitor's band limited RMS segments perform better than OSEMs in identifying it.}
    \label{fig:osemtrans}
\end{figure}

We also examined the performance of BLRMS segments against OSEM time series on 3 days in O3a dominated by slow scattering noise. Fig. \ref{fig:osemtrans} shows the efficiency over dead time for Apr 13, 2019, May 13, 2019, and June 25, 2019. Efficiency is the fraction of filtered h(t) triggers that coincide with BLRMS segments. Deadtime refers to the total duration of the segments as a fraction of the total observing duration for that day. A large value of efficiency over deadtime is preferred as the goal is accurate identification of  noise. For all the three days, BLRMS segments register higher efficiency over deadtime than OSEMs scattering segments. The \textbf{gwdetchar-scattering} algorithm with just OSEMs as scattering witnesses, is designed to capture the motion of optics throughout the interferometer. 
As we have identified the likely location of scattering, we can use a more specific approach by employing the transmitted photodiode’s BLRMS segments to capture the noise.

Of the two separate noise couplings mentioned in Sec. \ref{m0r0_scattering}  and Sec. \ref{transmon_scattering}, the PUM stage OSEM is sensitive to only first of these, the ETM-AERM relative motion. And in principle we expect it to identify as much noise as the BLRMS segments. Following RC tracking however, the scattering impact due to ETM AERM relative motion has been reduced whilst having no effect on ETM TMS noise which the PUM stage is not sensitive to. Post RC tracking slow scattering noise can thus still be identified using the BLRMS segments.

\newpage
\section{Summary and discussion}\label{summary}
 Scattering noise affects the data quality of the Advanced LIGO detectors. Upconversion of the low-frequency noise due to large optic motion reduces the sensitivity of the detector in $10$ - $120$ Hz band. We use witnesses that identify times in the data when scattering noise is present as well as, when possible, identifying and eliminating the scatter mechanism in the instrument detector itself. 

We analyzed light scattering in LLO during O3 and we found the presence of two different populations of scattering noise, slow scattering, and fast scattering. We investigated slow scattering that appears with a typical arch shape in the time-frequency representation and we found two different paths through which this noise couples to the detector simultaneously. We were able to implement a solution for the louder noise coupling that resulted in a substantial reduction of the noise and we discussed possible remedies for the second one. One of these solutions, the TMS feed forward, we plan to implement in O4. 
In order to identify the times when this noise is present in the gravitational wave channel, we suggested using the band-limited time-series data of an auxiliary channel. This channel, monitors the light transmitted through the end test mass and we showed it identifies a larger subset of scattering triggers as compared to other scattering witnesses.

High Q resonances found at the corner and end stations at LLO could be contributing to fast scattering. The ongoing investigation suggests that damping the motion of some optical components at these stations would likely mitigate the rate of fast scattering.
\par

\section{Acknowledgements} LIGO was constructed by the California Institute of Technology and Massachusetts Institute of Technology with funding from the National Science Foundation and operates under Cooperative Agreement No. PHY-1764464. Advanced LIGO was built under Grant No. PHY-0823459. We acknowledge support from the NSF grant PHY-1806656, PHY-1505779, and  PHY-1912604. We also acknowledge the discussions with the members of Detector Characterization Group of the LIGO Scientific Collaboration. For this paper, we use the data from the Advanced LIGO detectors and we used the LIGO computing clusters to perform the analysis and calculations.



%% References


%\bibliography{sample.bib}



\newpage
\section*{References}
\bibliographystyle{unsrt}
\bibliography{sample.bib}
%\begin{thebibliography}{00}


%\bibitem{1}

%\end{thebibliography}
%\section{Referencess}
%\begin{itemize}
%    \item [1] Advanced Ligo LIGO-P1400177-v5
%    \item [2] Acernese F et al. (Virgo Collaboration) 015 %Class. Quant. Grav. 32 024001 (Preprint %arXiv:1408.3978)
%    \item [3]  Dooley K L et al. 2016 Class. Quant. Grav. %33 075009 (Preprint arXiv:1510.00317) URL %https://doi.org/10.1088/0264- 9381/33/7/075009
%    \item [4] AkutsuTetal.(KAGRA)2017ThestatusofKAGRAundeg%roundcryogenicgravitationalwave telescope 15th %International Conference on Topics in Astroparticle %and nderground Physics (TAUP 2017) Sudbury, Ontario, %Canada, uly 24-28, 2017 (Preprint arXiv:1710.04823)
%    
%\item [5] B. J. Meers, Phys. Rev. D 38, 2317 (1988).
%\item [6] GWTC-1: A Gravitational-Wave Transient Catalog %of Compact Binary Mergers Observed by LIGO and Virgo %during the First and Second Observing Runs. %arXiv:1811.12907
%\item [7] On the Progentor of Neutron Star Merger %GW170817.
%arxiv.1710.05838
%\item [8] Prospects for Observing and Localizing %Gravitational-Wave Transients with Advanced LIGO, dvanced %%Virgo and KAGRA arXiv:1304.0670v9
%\item [9] Detector Characterization for Advanced Ligo, %Thesis Thomas J Massinger.
%\item [9] Noise from scattered light in Virgo's second %science run data. T Accadia et al 2010 Class. Quantum %Grav. 27 194011
%\item [10] https://alog.ligo-la.caltech.edu/aLOG/index.ph%%?callRep=45053
%\end{itemize}

\end{document}